\documentclass[apj,onecolumn]{emulateapj}

\newcommand{\FF}{\mbox{\boldmath$F$}}
\newcommand{\LL}{\mbox{\boldmath$L$}}
\newcommand{\VV}{\mbox{\boldmath$V$}}

\newcommand{\xx}{\mbox{\boldmath$x$}}
\newcommand{\yy}{\mbox{\boldmath$y$}}
\newcommand{\zz}{\mbox{\boldmath$z$}}
\newcommand{\ee}{\mbox{\boldmath$e$}}
\newcommand{\rr}{\mbox{\boldmath$r$}}
\newcommand{\hcur}{\mbox{${\mathcal{H}}$}}
\usepackage{epsfig}
\newif\ifAMStwofonts
\begin{document}
\title{Binaries traveling through a gaseous medium: dynamical drag forces and internal torques}
\shorttitle{Binaries in a gaseous medium}

\author
{F.~J.~S\'anchez-Salcedo\thanks{E-mail:jsanchez@astro.unam.mx}}
\affil{Instituto de Astronom\'{\i}a, Universidad Nacional Aut\'onoma
de M\'exico, Ciudad Universitaria, Apt.~Postal 70
264, \\ C.P. 04510, Mexico City, Mexico}

\and

\author{Raul O. Chametla}

\affil{Escuela Superior de F\'{\i}sica y Matem\'{a}ticas, Instituto Polit\'{e}cnico Nacional, UP Adolfo
L\'opez Mateos, Mexico City, Mexico}

\shortauthors{S\'anchez-Salcedo \& Chametla}



\begin{abstract}

Using time-dependent linear theory,
we investigate the morphology of the gravitational wake induced by a binary, whose
center of mass moves at velocity $\VV_{\rm cm}$
against a uniform background of gas.  For simplicity, we assume that the binary's
components are on circular orbits about their common center of mass. 
The consequences of dynamical friction is twofold.
First, gas dynamical friction may drag the binary's center
of mass and cause the binary to migrate. 
Second, drag forces also induce a braking torque, which causes the
orbits of the binary components to shrink.  We compute the drag forces acting
on one component of the binary due to the gravitational interaction with its own wake.
We show that the dynamical friction force responsible to decelerate the binary's center 
of mass is smaller than it is in the point-mass case because of the loss of gravitational
focusing.
We show that the braking internal torque depends on the Mach numbers of each 
binary component about their center of mass,
and also on the Mach number of the center of mass of the binary. 
In general, the internal torque decreases with increasing the velocity of 
the binary relative to the ambient gas cloud. However, this is not always the case.
We also mention the relevance of our results on the period distribution of binaries.


\end{abstract}


\section{Introduction}

Stars and black holes can accrete gas when they move through a gaseous
medium (Bondi \& Hoyle 1944). Gas accretion may shape the 
observed initial mass function of stars in young star-forming clusters 
(e.g., Bonnell et al. 2001; Maschberger et al. 2014). In addition to the aerodynamic
drag force $\dot{M}V$ due to the accretion of gas,
gravitational objects moving in a gas may also feel a drag or dynamical friction 
due to the gravitational interaction with its own-induced wake.
There is a variety of astrophysical systems where gas dynamical friction
plays an important role; e.g., gas dynamical friction is relevant to understand the orbital
decay of common-envelope binary stars 
(Taam \& Sandquist 2000; Nordhaus \& Blackman 2006; Ricker \& Taam 2008),
the orbital decay of giant clumps in high-redshift galaxies (Immeli et al. 2004;
Bournaud et al. 2007) or the shrinkage of the orbit of supermassive black holes 
in merging galaxies (e.g., Armitage \& Natarajan 2002).
 Dynamical friction may affect the
stellar dynamics in star clusters still embedded in their parent gas cloud, 
and/or a nuclear star cluster subject to major gas inflows (e.g., Davies et al. 2011).
For instance,
Chavarr\'{\i}a et al. (2010) suggested that the mass segregation observed in two
clusters of young stellar objects associated with massive star-forming regions in
the Norma Spiral Arm could be explained by the migration of high-mass 
stars toward the center due to gaseous gravitational drag. Leigh et al. (2014) found that
for the galactic nuclei and young star-forming regions considered in their study,
the rate of mass segregation due to gas dynamical friction and accretion tends to
be comparable to, albeit slightly smaller than, the rates from stellar two-body relaxation.
Thus, gas damping can accelerate the rate of mass segregation.

In the work, we consider
the problem of the orbital evolution of binaries embedded in gaseous media, e.g., 
in a cloud\footnote{The evolution of binaries in 
stellar clusters due to the dynamical interactions with other
stars has received considerable attention (e.g., Heggie 1975; Hills 1975; Hut 1983; 
Hut et al. 1992;  Sigurdsson \& Phinney 1993;
Downing et al. 2010).}. 
Dynamical friction will lead
to a migration of the binary to the center of the cloud, producing
mass segregation. In addition, 
dynamical interaction with the surrounding gas may produce a braking torque,
which can lead to the extraction
of energy and angular momentum from the binary (Kim et al. 2008;
Stahler 2010). This torque tends to shrink the separation of the binary components,
decreasing its orbital period around the center of mass of the binary.
Korntreff et al. (2012) argued that during the time that the cluster
is embedded in its natal gas ($\sim 1$ Myr), the gas-induced orbital decay can reshape the period
distribution in close systems (separations $\leq 30$ AU). Moreover, they suggested
that gas damping can catalyse the coalescence of star binaries, reducing
the binary frequency.

One would expect that, since binary systems are, on average, more massive than single 
stars, they should present a more clear signature of mass segregation. Contrary to these
expectations, de Grijs et al. (2013) and Li et al. (2013) showed evidence that the fraction of F-star
binary systems in NGC 1818 increases with increasing distance from the cluster center.
This could be indicative that either the dynamical friction is reduced in binary
systems or, more likely, that they are observing hard binary systems (those with relatively
high binding energies compared to the kinetic energy of bulk stellar population) 
that have survived and could have been hardened by dynamical encounters and by
gas dynamical friction. Geller et al. (2013) found that, depending on the dynamical age
of the cluster, the radial binary frequency distribution
can either increase or decrease moving out radially from the cluster center, 
due to the combined effects of binary disruption and mass segregation.

Gas dynamical friction is also an important ingredient to understand the orbital decay of supermassive black holes 
at the center of galaxies (e.g., Armitage \& Natarajan 2002; see Colpi 2014 for a review). 
It is believed that all galaxies with stellar
spheroids possess a supermassive black hole. Thus, a merger of two galaxies
leads to the formation of a supermassive black hole binary. If the timescale
for coalescence of the black holes in the binary is longer than the 
time for succesive major mergers, then the galaxy may undergo a
subsequent merger when the binary is still in place (Hoffman \& Loeb 2007; 
Amaro-Seoane et al. 2010; Kulbarni \& Loeb 2012). 
When a galaxy with a binary black hole merges with other galaxy that contains
a central black hole,
their black holes sink to the center of the merger product by dynamical friction, forming
a triple black hole system.
Therefore, it is interesting to understand the orbital decay of the center-of-mass of
a binary black hole, due to dynamical
friction,  in the aftermath of a gas-rich merger. In addition to the orbital decay towards the
center of the remnant, the binary is subject to loss of angular momentum about
its center of mass, which means the separation between the
components of the binary shrinks.
Note that black hole binaries are expected to be surrounded by a rotating stellar and gaseous 
disk, resulting from the merger of the two parent disks. The associated drag depends on 
the differential rotation of the disk, its self-gravity and on the presence of resonances. 
Therefore, models that assume that the background is initially uniform and static cannot be applied
to study the orbital decay of black hole binaries in the center of galaxies.

The orbital decay of a {\it single} perturber orbiting a gaseous cloud
was studied numerically by S\'anchez-Salcedo \& Brandenburg (2001)
and by Escala et al. (2004). S\'anchez-Salcedo \& Brandenburg (2001) found
that an extension of the analytical formula derived by Ostriker (1999) for perturbers
in rectilinear orbit and constant velocity in homogeneous media,  is very successful in describing
the orbital evolution. Escala et al. (2004) provided
a fitting formula for the decay timescale of a massive perturber in
a self-gravitating isothermal sphere. Cant\'o et al. (2013) derived a formula for the
gravitational drag on a body moving in a vertically stratified medium of gas.

Kim, Kim \& S\'anchez-Salcedo (2008) considered an embedded-gas
binary and derived the braking forces on each particle assuming that the two components
of the binary are on circular orbits. 
Stahler (2010) studied the same problem but used a different approach to
estimate the angular momentum exchange between the binary and the surrounding gas.
In both calculations, it was assumed that the center-of-mass of the
binary is at rest relative to the gaseous medium. 
The goal of this paper is to derive the drag forces and internal torques acting on a binary system
in a more general situation in which the binary system is moving relative to the 
surrounding medium. We will see that the morphology of the wake, the drag force and the internal
torques are modified, in a subtle manner, when the additional component of the center-of-mass
velocity is included.

The paper is organized as follows. In Section \ref{sec:formulation}, we present the mathematical framework 
to derive the gravitational wake excited by a binary system, using time-dependent linear
perturbation theory.   Estimates of the length scale at which the response of the gas departs
from linearity are also discussed in Section \ref{sec:formulation}. 
Since in linear theory the wake created by two perturbers is given by the simple
superposition of the wakes of both perturbers,
Section \ref{sec:wake}
describes the morphology of the resulting wake created by just one component of
the binary. In Section \ref{sec:drag}, we evaluate the drag foce on the perturber due to
its own wake.  In Section \ref{sec:equalmass} we provide
estimates of the drag forces and internal torques in equal-mass
binaries and outline how to determine the binary orbital evolution. 
Section \ref{sec:conclusions} contains our conclusions.

\section{Formulation of the problem, relevant scales and linear theory}
\label{sec:formulation}
Consider a self-gravitating bound system whose center of mass
moves at a velocity $\VV_{\rm cm}$ relative to the gaseous background.
The system could be a globular cluster orbiting the gaseous
halo of a protogalaxy, a star cluster in the gaseous disk of a galaxy,
or a binary star embedded in its progenitor cloud. To make the presentation
readeable, we will consider the context of a binary system, but
our approach and results are also valid and applicable in many other 
astrophysical scenarios.

\subsection{The orbit of one component of the binary}
\label{sec:orbit}

We will compute the response of the gaseous medium to the gravitational
potential created by a binary system that travels in a linear trajectory through an initially homogeneous
medium with density and sound speed at infinity $\rho_{\infty}$ and $c_{\infty}$, respectively.
As usual in studies of dynamical friction, we will assume that the orbital parameters of the binary 
are constant over time, i.e. we ignore the effect of the drag force on the orbit.
Without loss of generality, we assume that the center
of mass of the binary moves along the $z$-axis with velocity $V_{\rm cm}$,
that is, $\VV_{\rm cm}=V_{\rm cm}\hat{\zz}$.
In the binary frame, the two particles move in a plane; we will refer to it as the orbital plane.

Denote $i$ as the inclination angle of the orbit (where $i$ is assumed to lie between
$0$ and $\pi/2$), i.e. the angle between the vector 
perpendicular to the orbital plane
and $\VV_{\rm cm}$. 
When $i=0$, the orbital axis is aligned
with the velocity of the center of mass and, thus, the binary moves face-on
relative to the ambient medium. On the other hand, $i=\pi/2$ corresponds to
an edge-on motion.
Finally, we define a unit vector $\ee_{a}$ along the intersection of the orbital plane
with the reference plane $(\hat{\xx},\hat{\yy})$. We can always choose a
system of reference where $\ee_{a}$ is along the $x$-axis (that is,
the longitude of the ascending node is zero).

For simplicity, we consider  the case where the orbit of the
binary is circular,
and focus on the wake created by one of the components of the
binary, with mass $M$ (hereafter the perturber).
We let $R_{p}$ denote the radius of its orbit around the center of mass of the binary. 
In the frame of reference described above, the orbit is given by
\begin{equation}
x_{p}(t)=R_{p}\cos\Omega t, 
\label{eq:xp}
\end{equation}
\begin{equation}
y_{p}(t)=R_{p}\cos i\sin \Omega t, 
\end{equation}
\begin{equation}
z_{p}(t)=R_{p}\sin i\sin\Omega t+V_{\rm cm}t,
\label{eq:zp}
\end{equation}
where we have assumed that, at $t=0$, the perturber is at $x=R_{p}$, $y=0$ and $z=0$ and,
in addition, for convention,  $\Omega \geq 0$ and $V_{\rm cm}\geq 0$.

\begin{figure}
\epsfig{file=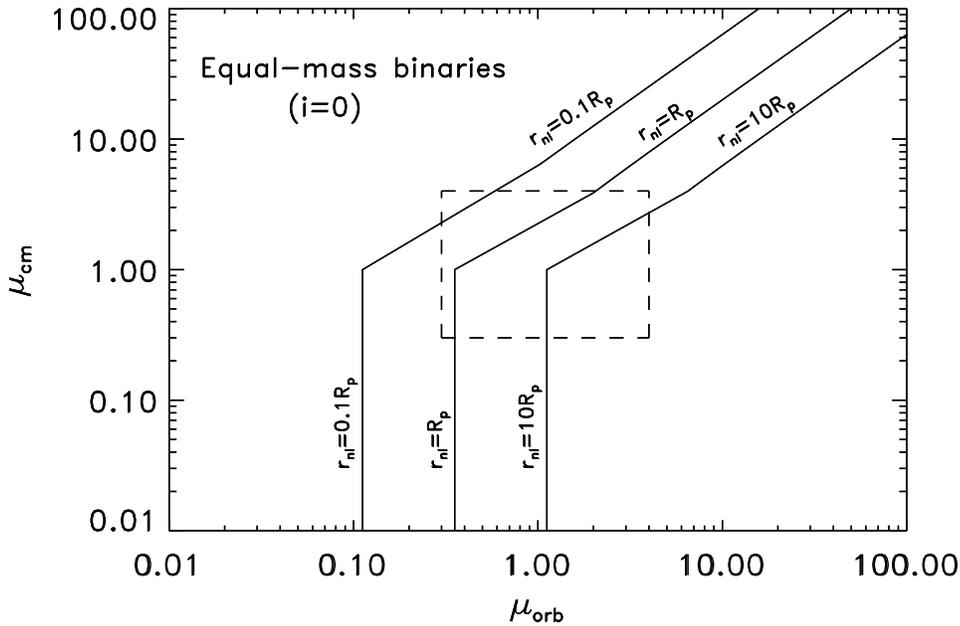,angle=0,width=13.0cm}
  \caption{The contour lines show $r_{\rm nl}$, which is defined as the characteristic radius where the dynamics
becomes nonlinear, in the plane $(\mu_{\rm orb},\mu_{\rm cm})$, for an equal-mass binary
with $i=0$. Each curve indicates
the different combinations of $\mu_{\rm orb}$ and $\mu_{\rm cm}$ that give the same
$r_{\rm nl}$. The box indicates the parameter space explored in this paper. }
\label{fig:parameter_space}
\vskip 0.75cm
\end{figure}

The gas will respond to the external gravitational potential created by the
perturber, $\Phi_{\rm ext}$, which obeys the Poisson equation $\nabla^{2}\Phi_{\rm ext}=4\pi G\rho_{\rm ext}$,
where $\rho_{\rm ext}$ is the density profile of the perturber. In our case:
\begin{equation}
\rho_{\rm ext}(\rr,t)=M\delta(x-x_{p}(t))\delta(y-y_{p}(t))\delta(z-z_{p}(t)) \hcur(t),
\label{eq:rho_ext}
\end{equation}
where we are assuming that the perturber is a point-mass.
Here $\hcur$ is an arbitrary function of $t$ which may be used to describe how
the perturber is introduced in the medium; $\hcur=1$ corresponds to the
stationary state, where the perturber is present since $t\rightarrow -\infty$. 
If $\hcur$ is the Heaviside function, $\hcur(t)=\Theta(t)$
the perturber is dropped suddently at $t=0$, whereas if $\hcur(t)=(1-\exp[-t/\tau])\Theta(t)$, the
perturber is inserted adiabatically when $\tau$ is large enough. Ostriker (1999) 
noticed that for subsonic perturbers, the stationary solution does not capture all
the physics of the problem.

It is useful to define the Mach number of the center of mass $\mu_{\rm cm}\equiv V_{\rm cm}/c_{\infty}$
and the orbital Mach number defined as $\mu_{\rm orb}\equiv V_{\rm orb}/c_{\infty}$, where 
$V_{\rm orb}\equiv\Omega R_{p}$. A model can thus be specified with three dimensionless
parameters $(i, \mu_{\rm cm}, \mu_{\rm orb})$. 

\subsection{Relevant length scales and applicability of the linear theory}
\label{sec:scales}

Discussions are more simple in the case where $i=0$ because the
modulus of the velocity of each perturber is constant. For $i>0$, the relative 
velocity between one perturber and the ambient medium
depends on the position along the orbit, which complicates the treatment. Therefore,
we will focus on the face-on case ($i=0$).

Consider for a moment the wake created by just one object of mass $M$ that is forced to move in a
helical motion without any other gravitational companion.
If the perturber is a point-mass, there exists a neighbourhood around the
body where the response of the gas becomes nonlinear. Let us denote $r_{\rm nl}$
the radius of this region.
When the guiding center moves subsonically, $\mu_{\rm cm}<1$,
the response of the gas is nonlinear within the Bondi radius, $r_{B}$, defined as the
distance at which the gravitational potential is comparable to the gas pressure, 
that is $r_{\rm nl}=r_{B}\equiv GM/c_{\infty}^{2}$. 
Now suppose that the motion of the guiding center is supersonic:
$\mu_{\rm cm}>1$.
It is convenient to define the gravitational radius $r_{gc}\equiv GM/c_{\infty}^{2}(1+\mu_{\rm cm}^{2})$.
If $r_{gc}\ll R_{p}$, then curvature effects of the orbit are not
important and,
therefore,  the response becomes nonlinear within the accretion radius\footnote{The accretion 
radius is defined as the radius at which the blending
of streamlines is important.}, 
$r_{acc}=GM/c_{\infty}^{2}(1+\mu_{\rm cm}^{2}+\mu_{\rm orb}^{2})$, which corresponds
to the accretion radius of a body in rectilinear orbit with effective Mach number 
$(\mu_{\rm cm}^{2}+\mu_{\rm orb}^{2})^{1/2}$. 
If, on the other hand, $R_{p}\ll r_{gc}$, the orbit is so close that its size  
and the orbital velocity $V_{\rm orb}$ are both
irrelevant. Under this condition,  the problem is nonlinear at  distances $\lesssim r_{gc}$ (i.e.,
$r_{\rm nl}=r_{gc}$). Since $r_{acc}$ and $r_{gc}$ are both proportional to $M$, $r_{\rm nl}$
also depends on $M$. Hence, $r_{\rm nl}$ can be arbitrarily small by decreasing the value of $M$.

In the presence of a companion, it is necessary to compute $r_{\rm nl}$ for each constituent of the binary,
say $r_{{\rm nl},1}$ and $r_{{\rm nl},2}$.
If $r_{{\rm nl},1}+r_{{\rm nl},2}<a$, where $a$ is the separation of the binary, 
the nonlinear parts of the wakes do not intersect.  Thus, the fluid is nonlinear in regions
near each perturber with characteristic radii $r_{\rm nl,1}$ and $r_{\rm nl,2}$. For unequal-mass
binaries with $M_{1}\gg M_{2}$, the condition $r_{{\rm nl},1}+r_{{\rm nl},2}<a$ is satisfied either
when $\mu_{\rm orb}<1$ and $\mu_{\rm cm}<1$,
or when $\mu_{\rm cm}^{2}>{\rm max}\{\mu_{\rm orb}^{2}-1,1\}$.

\begin{figure}
\epsfig{file=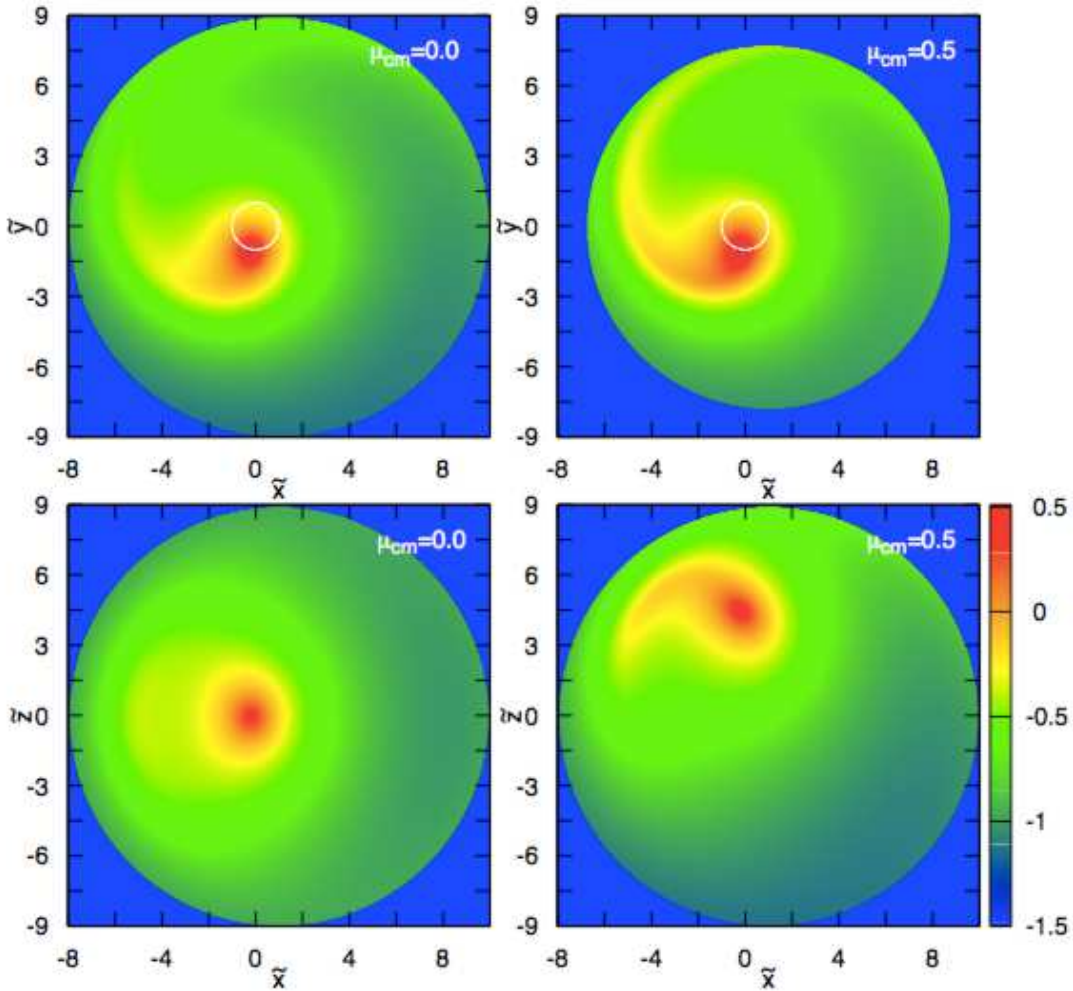,angle=0,width=18.5cm}
  \caption{Color maps of the perturbed density ${\mathcal{D}}$ at $\tilde{t}=9$
generated by a body 
in face-on motion with $\mu_{\rm orb}=0.5$ and
two different values of $\mu_{\rm cm}$. The top panels correspond to cutoffs through the $(x,y)$-plane
at $\tilde{z}=0$ (left) and $\tilde{z}=4.5$ (right). These planes contain the perturber which can be distinguished
as a very tiny black point. The white circles indicate the orbit of the perturber in each plane.
The bottom panels show cutoffs through the plane $\tilde{y}=-0.5$.
The perturber is located at $\tilde{r}_{p}=( -0.21,-0.98 , 0)$ when $\mu_{\rm cm}=0$, and at $(-0.21,-0.98,4.5)$ when $\mu_{\rm cm}=0.5$.
 }
\label{fig:mu_subsonic}
\end{figure}

\begin{figure}
\epsfig{file=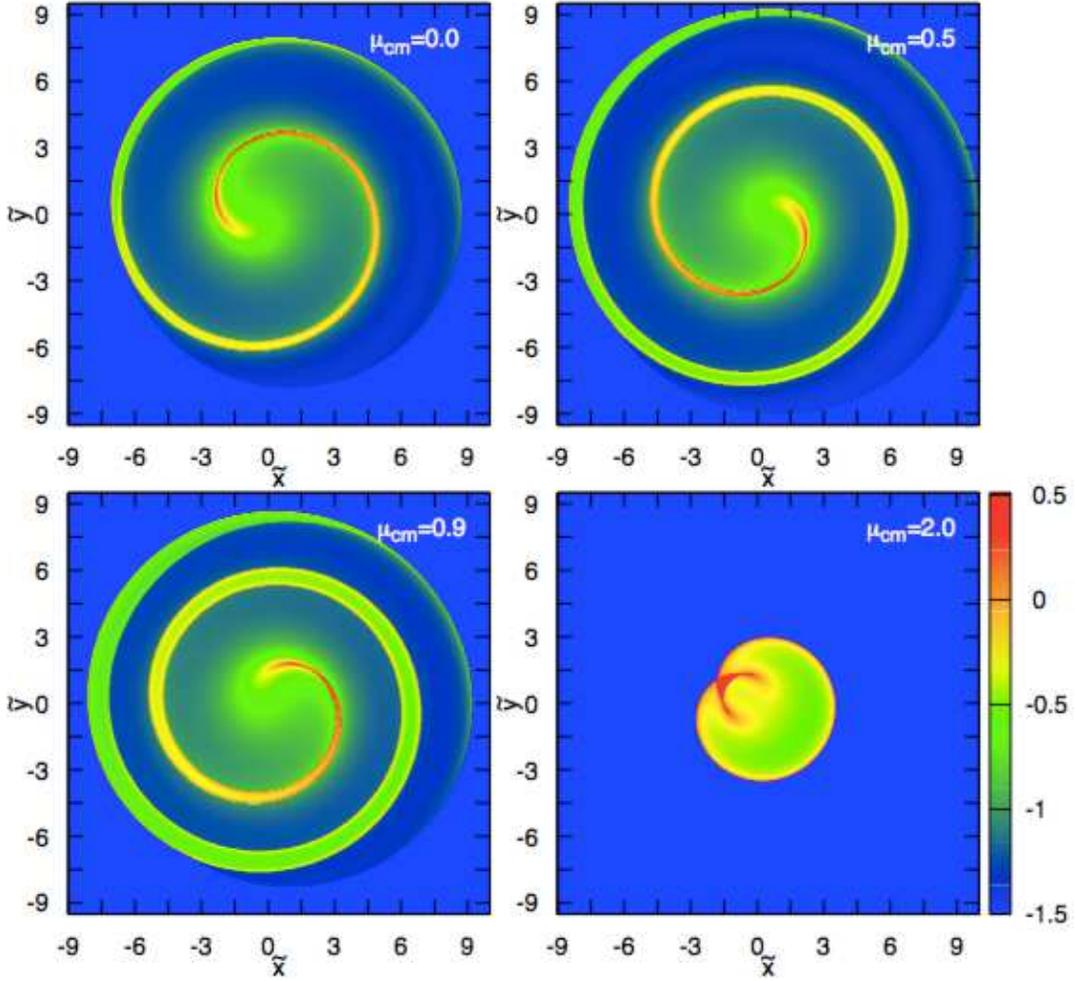,angle=0,width=18.5cm}
 \caption{Density enhancement ${\mathcal{D}}$ created by a face-on perturber with
$\mu_{\rm orb}=2$, for different values of $\mu_{\rm cm}$,
along a cutoff through the
$(x,y)$-plane at $\tilde{z}=\tilde{z}_{p}-4.5$, i.e., through planes situated
at a vertical distance $4.5R_{p}$ behind the perturber. The time selected was $\tilde{t}=9$ and thereby 
$\tilde{z}_{p}=\mu_{\rm cm}t=9\mu_{\rm cm}$. }
\label{fig:muorb2_horizontal}
\end{figure}

\begin{figure}
\epsfig{file=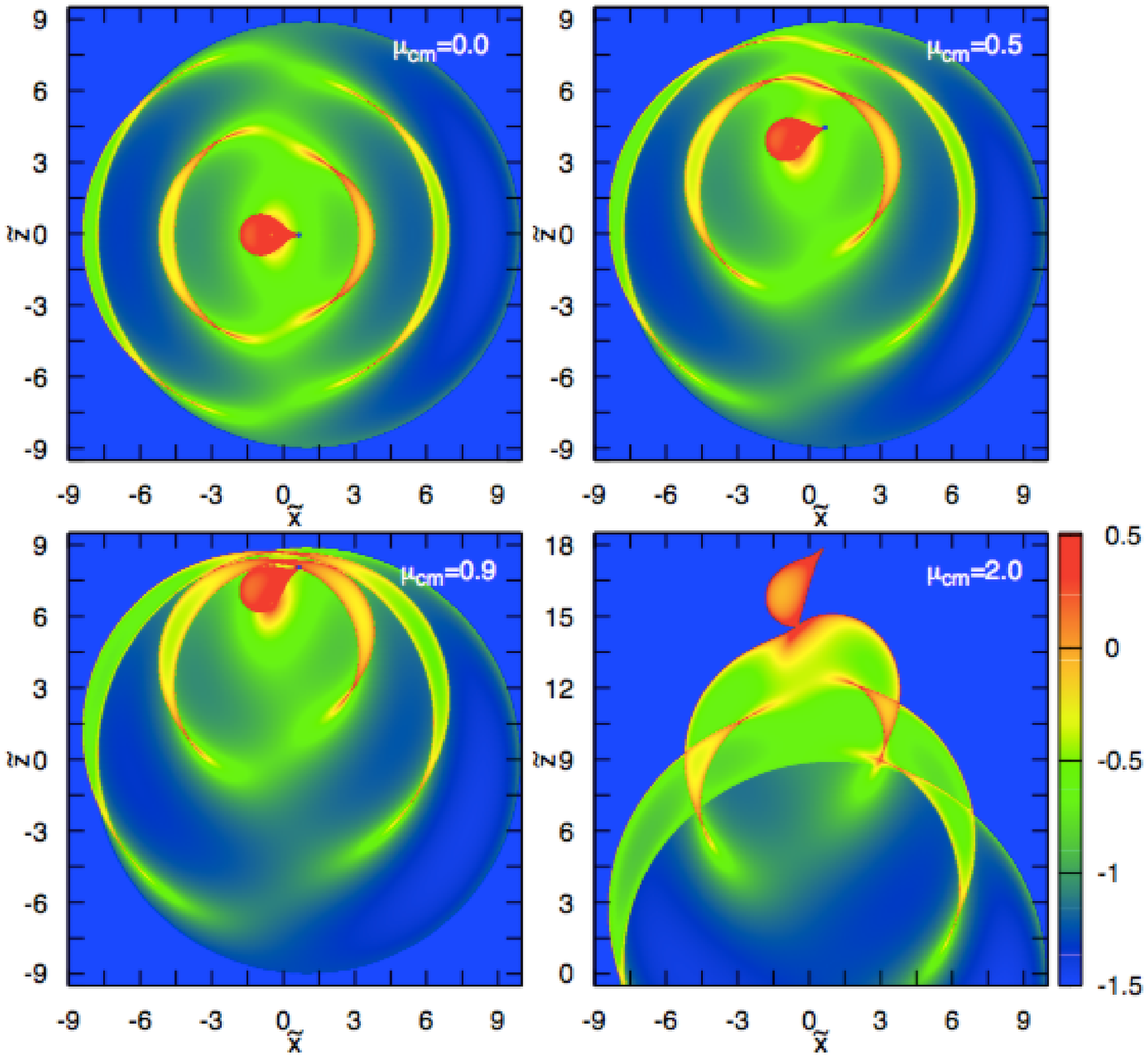,angle=0,width=18.5cm}
\caption{Same as Figure \ref{fig:muorb2_horizontal} but along a cutoff 
through the vertical plane $\tilde{y}=-0.8$, which passes through
the location of the perturber at that time.}
\label{fig:muorb2_vertical}
\end{figure}

Now consider an equal-mass binary, each component having mass $M$. 
The value of $R_{p}$ and the orbital velocity are not
independent but related through $V_{\rm orb}^{2}=GM/(4R_{p})$. Therefore,
$R_{p}=GM/(4V_{\rm orb}^{2})$. The Bondi radius of the binary, which has total
mass $2M$, is  $r_{B}=G(2M)/c_{\infty}^{2}=8\mu_{\rm orb}^{2}R_{p}$. Consequently,
if $\mu_{\rm orb}>0.35$, then $r_{B}>R_{p}$. This means that if $\mu_{\rm cm}<1$ and
$\mu_{\rm orb}>0.35$, the flow is in the nonlinear regime within a sphere of radius 
$\sim 8\mu_{\rm orb}^{2}R_{p}$ around the binary.

If, on the other hand, the center of mass of an equal-mass binary
moves supersonically at, say $\mu_{\rm cm}=5$, then $r_{gc}=GM/c_{\infty}^{2}(1+\mu_{\rm cm}^{2})\simeq 
(1/6)\mu_{\rm orb}^{2}R_{p}$. As a result, for $\mu_{\rm orb}\lesssim 1$, the flow becomes nonlinear
in regions of size smaller than $R_{p}$ near each component of the binary.

Following these arguments, we have constructed Figure \ref{fig:parameter_space},
which shows contour lines
of $r_{\rm nl}$ in the plane $(\mu_{\rm orb}, \mu_{\rm cm})$ for an equal-mass binary.
In Section \ref{sec:wake}, we calculate the wake for $\mu_{\rm orb}$ and $\mu_{\rm cm}$
ranging between $0.3$ and $4$. As we see in Figure \ref{fig:parameter_space}, $r_{\rm nl}$
may be as small as $0.1R_{p}$ in the limit case that 
$\mu_{\rm orb}\simeq 0.3$ and $\mu_{\rm cm}\simeq 4$. Therefore, {\it in equal-mass binaries}
and for the parameters selected in this investigation,
the linear approximation theory definitely breaks down at
distances less than $\sim 0.1 R_{p}$ from the perturber, or even at larger distances for other 
combinations of $(\mu_{\rm orb}, \mu_{\rm cm})$.  

\begin{figure}
\epsfig{file=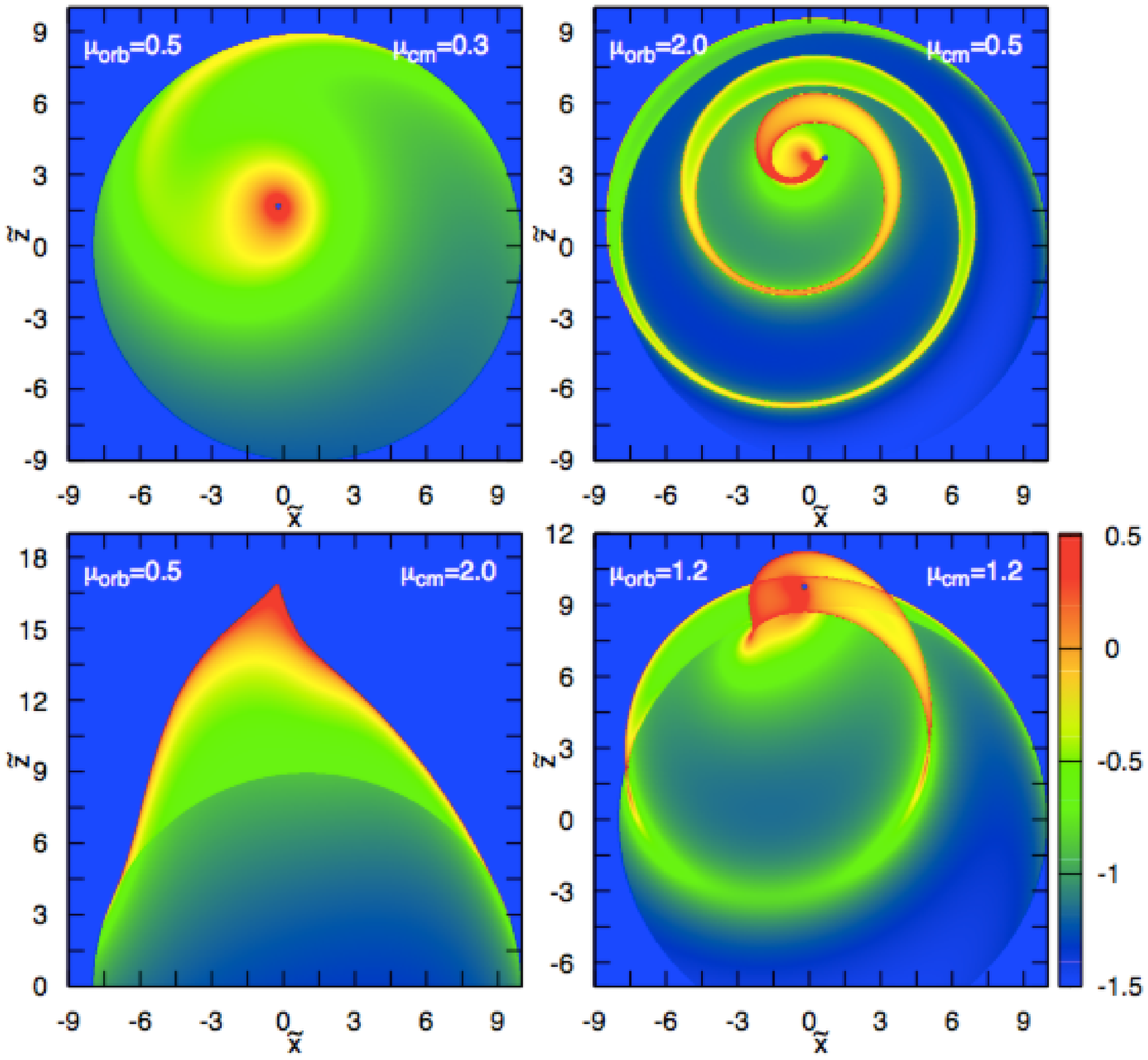,angle=0,width=18.5cm}
\caption{Distribution of the perturbed density ${\mathcal{D}}$ through the plane $y=0$
for edge-on perturbers, at $\tilde{t}=9$. The corresponding values of $\mu_{\rm cm}$ and $\mu_{\rm orb}$
are quoted at the right and left corners in each panel. Note that the trajectory of
the perturber always lies in the plane $y=0$.
 }
\label{fig:D_edgeon}
\end{figure}

\subsection{The gravitational wake in linear theory}
In the previous section (\S \ref{sec:scales}), we have estimated the characteristic scales
at which the flow becomes nonlinear. 
Far enough from the gravitational object, the perturbation is so small that it
can be treated in the linear approximation.
Combining the linearized equation of
continuity and the linearized equation of motion, it is straightforward to find the
differential equation for $\alpha(\rr,t)\equiv (\rho-\rho_{\infty})/\rho_{\infty}$, where
$\rho(\rr,t)$ is the gas density of the medium. The governing equation for $\alpha(\rr,t)$ reads
\begin{equation}
\nabla^{2}\alpha-\frac{1}{c_{\infty}^{2}}\frac{\partial^{2}\alpha}{\partial t^{2}}=
-\frac{4\pi G}{c_{\infty}^{2}}\rho_{\rm ext}(\rr,t),
\label{eq:inhomogeneous_equation}
\end{equation}
which corresponds to the inhomogeneous wave equation (e.g., Ostriker 1999).
The formal solution of this equation is given by:
\begin{equation}
\alpha(\rr,t)=\frac{G}{c_{\infty}^{2}}\int\int d^{3}\rr' dt' \frac{\delta[t'-t+|\rr-\rr'|/c_{\infty}]
\rho_{\rm ext}(\rr',t')}{|\rr-\rr'|}.
\label{eq:formal_solution}
\end{equation}
For a point-mass perturber in a straight-line orbit (i.e. $\Omega=0$), Equation (\ref{eq:formal_solution})
can be integrated analytically (Ostriker 1999). Here we will consider the case where the
perturber follows 
 a circular orbit around a guiding center (the binary's barycenter) which in turn moves on 
a straight-line trajectory (Eqs. \ref{eq:xp}-\ref{eq:zp}).
Substituting the expression for $\rho_{\rm ext}(\rr,t)$
given in Eq. (\ref{eq:rho_ext}) into Equation
(\ref{eq:formal_solution}) and integrating over $x'$, $y'$ and $z'$, Equation (\ref{eq:formal_solution})
can be reduced to 
\begin{equation}
\alpha(\rr,t)=\frac{GM}{c_{\infty}^{2}R_{p}} \int \frac{\delta[\varphi-\Omega t+\mu_{\rm orb}\tilde{d}(\varphi;\rr)]\hcur(\varphi/\Omega)}{\tilde{d}(\varphi;\rr)} d\varphi,
\label{eq:alpha_int}
\end{equation}
where we have used the new variable $\varphi\equiv \Omega t'$ 
(note that $\Omega\neq 0$) and 
\begin{eqnarray}
\tilde{d}(\varphi;\rr)=\left[\left(\tilde{x}-\cos\varphi\right)^{2}+\left(\tilde{y}-\cos i\sin \varphi\right)^{2} 
+\left(\tilde{z}-\sin i\sin\varphi -\lambda\varphi\right)^{2}\right]^{1/2},
\end{eqnarray}
where $\lambda\equiv \mu_{\rm cm}/\mu_{\rm orb}$.
Remind that $\mu_{\rm orb}=\Omega R_{p}/c_{\infty}$.
 In the above equation (\ref{eq:alpha_int}) and throughout the paper, the tilde over one variable is used to indicate 
dimensionless quantities. The length scales are made dimensionless with $R_{p}$. Specifically, 
$\tilde{d}\equiv d/R_{p}$, $\tilde{x}\equiv x/R_{p}$ and so on.

\begin{figure}
\epsfig{file=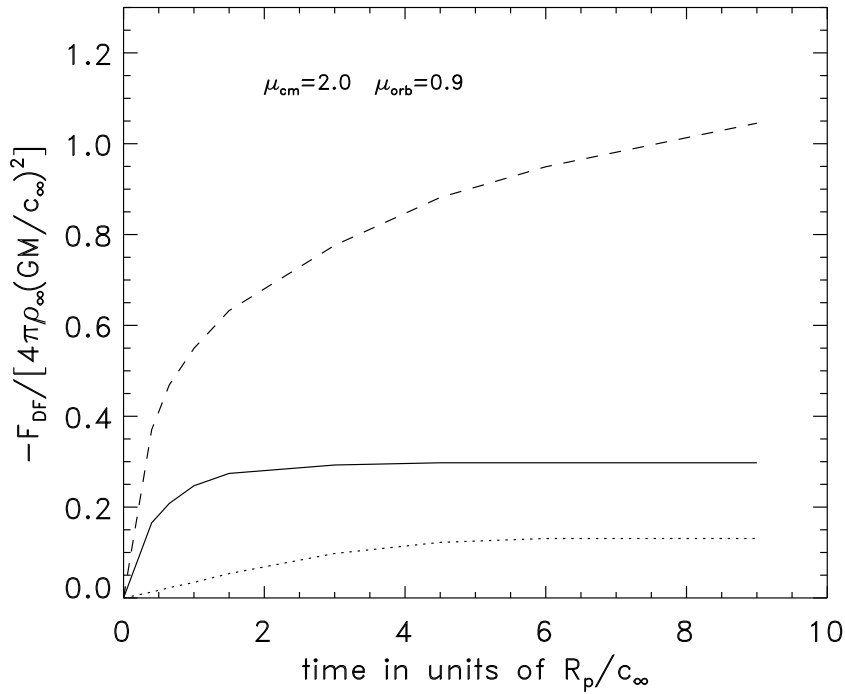,angle=0,width=12.5cm}
  \caption{Temporal evolution of the $z$-component (dashed line), azimuthal component (solid line)
and radial component (dotted line) of the dimensionless gravitational force for a single perturber
in a face-on orbit with $\mu_{\rm cm}=2$ and $\mu_{\rm orb}=0.9$. 
}
\label{fig:force_vs_time}
\vskip 0.75cm
\end{figure}

\begin{figure}
\includegraphics[width=130mm,height=140mm]{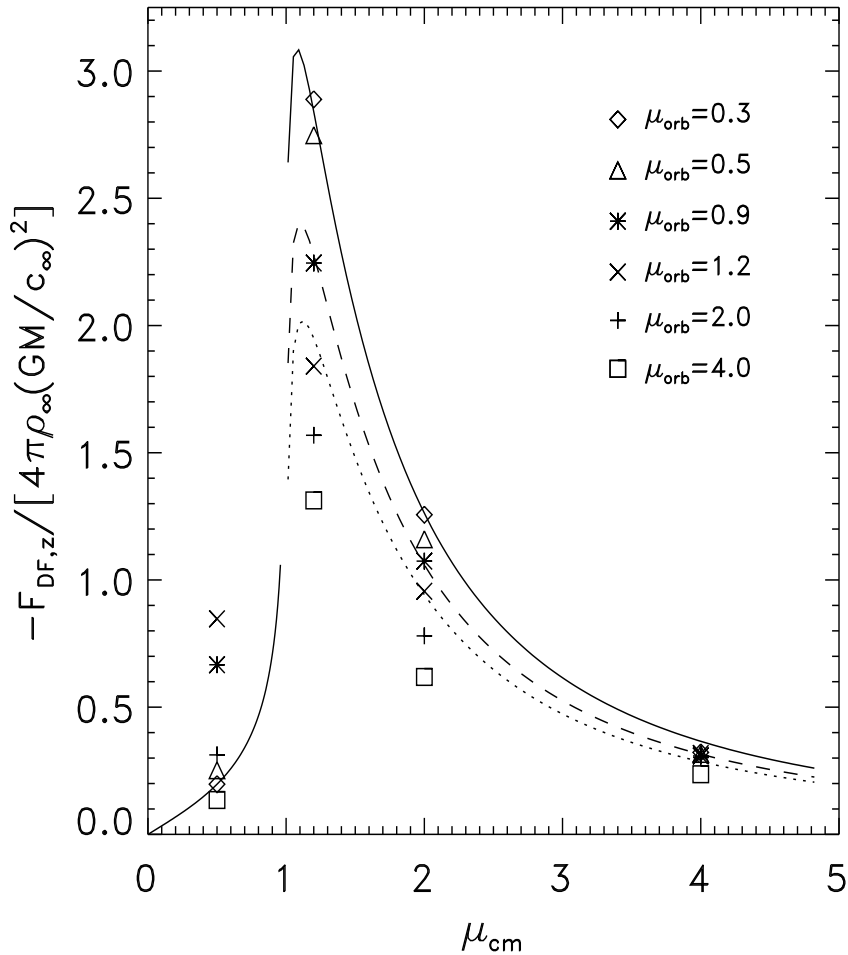}
  \caption{Vertical component (i.e. along the $z$-direction)
of the dimensionless dynamical friction force for a gravitational
object in the face-on case, as a function of $\mu_{\rm cm}$. The force was computed at $\tilde{t}=9$.
The  curves draw the drag force as predicted by Ostriker's formula (1999) for a body moving
at Mach number $\mu_{\rm cm}$ and adopting $r_{\rm inf}=r_{\rm min}$
(solid line), $r_{\rm inf}=2.25r_{\rm min}=0.225R_{p}$ (dashed line) and 
$r_{\rm inf}=3.6r_{\rm min}=0.36R_{p}$ (dotted line). 
 }
\label{fig:fz_vs_com_face}
\vskip 0.75cm
\end{figure}

In order to integrate Equation (\ref{eq:alpha_int}) over $\varphi$, we use the identity $\delta(g(\varphi))=\sum \delta(\varphi-\varphi_{j})/|g'(\varphi_{j})|$, where $\varphi_{j}$ are the roots of the function $g(\varphi)$. In our case,
$g(\varphi)=\varphi-\Omega t+\mu_{\rm orb}\tilde{d}$, and its derivative is
\begin{eqnarray}
&&g'\equiv \frac{dg}{d\varphi}=1+\frac{\mu_{\rm orb}}{\tilde{d}} \biggl( \tilde{x}\sin\varphi-
\tilde{y}\cos i\cos\varphi\\\nonumber
&&-\tilde{z}(\sin i\cos\varphi+\lambda)+\lambda\sin i (\sin\varphi+\varphi\cos\varphi)
+\lambda^{2}\varphi\biggr).
\end{eqnarray}
The $\varphi_{j}$-values are
the solutions of the following equation:
\begin{equation}
\mu_{\rm orb}\tilde{d}(\varphi;\rr)=-(\varphi-\Omega t).
\end{equation}
Evaluating the integral in Equation (\ref{eq:alpha_int})
and rearranging the terms, the perturbed density can be written as
\begin{equation}
\alpha (\rr, t)=\frac{GM}{c_{\infty}^{2}R_{p}}{\mathcal{D}}(\rr. t),
\end{equation}
with 
\begin{eqnarray}
{\mathcal{D}}(\rr,t)=\sum_{\varphi_{j}}\frac{\mu_{\rm orb}}{|(1-\mu_{\rm cm}^{2})\varphi_{j}-\Omega t-\mu_{\rm orb}^{2} 
[\tilde{x}\sin\varphi_{j}-\tilde{y}\cos i\cos\varphi_{j}-\tilde{z}(\sin i\cos\varphi_{j}+\lambda)+h(\varphi_{j})]|}
\hcur\left(\frac{\varphi_{j}}{\Omega}\right),
\label{eq:D_general}
\end{eqnarray}
where 
\begin{equation}
h(\varphi)=\lambda\sin i(\sin\varphi+\varphi\cos\varphi).
\end{equation}
If we use $R_{p}/c_{\infty}$ as the time unit, so that $\tilde{t}=t/(R_{p}/c_{\infty})$, then
$\Omega t=\Omega R_{p}\tilde{t}/c_{\infty}=\mu_{\rm orb}\tilde{t}$. Thus,
\begin{eqnarray}
{\mathcal{D}}(\rr,t)=\sum_{\varphi_{j}}\frac{\mu_{\rm orb}}{|(1-\mu_{\rm cm}^{2})\varphi_{j}-\mu_{\rm orb} \tilde{t}-\mu_{\rm orb}^{2} 
(\tilde{x}\sin\varphi_{j}-\tilde{y}\cos i\cos\varphi_{j}-\tilde{z}(\sin i\cos\varphi_{j}+\lambda)+h(\varphi_{j})|}
\hcur\left(\frac{\varphi_{j}}{\mu_{\rm orb}}\right),
\label{eqn:Dgen}
\end{eqnarray}
where the roots $\varphi_{j}$ are the solutions of the equation:
\begin{equation}
\mu_{\rm orb}\tilde{d}_{j}=-(\varphi_{j}-\mu_{\rm orb} \tilde{t}).
\label{eq:roots_gen}
\end{equation}
These expressions can be simplified for face-on orbits (see Appendix \ref{sec:app_faceon}),
and for edge-on orbits (see Appendix \ref{sec:app_edgeon}). Kim \& Kim (2007) studied the gravitational wake
created by a perturber on a pure circular orbit. Appendix \ref{sec:app_faceon} also shows that 
Kim \& Kim (2007) is recovered when $\mu_{\rm cm}=0$.

The method to find the overdensity $\mathcal{D}(\rr,t)$ is the same as described in
Kim \& Kim (2007). For the selected set of parameters $(i, \mu_{\rm cm}, \mu_{\rm orb})$, we
construct a grid. At a given time $\tilde{t}$, we find numerically the roots
of Equation (\ref{eq:roots_gen}) at each point of the grid $(\tilde{x}, \tilde{y}, \tilde{z})$,
using the Newton-Raphson bisection method. Appendix \ref{sec:app_location} 
establishes the procedure to find
the interval where the roots are located. Once the roots are found,
we evaluate Equation (\ref{eqn:Dgen}) to find $\mathcal{D}(\rr, t)$.
One of the advantages of the semi-analytical approach is that $\mathcal{D}$ can be derived
at the desired degree of accuracy. In addition, we can compute $\mathcal{D}$ at a time $\tilde{t}$ without
the need to follow its evolution at all intermediate timesteps, as occurs when solving numerically a differential
equation.  Since the problem is transformed into an algebraic one, the density can be computed
pixel-by-pixel, allowing to distribute the work in different processors or machines and thereby
achieving very high-resolution in modest desktop computers.

We must stress again that the linear approximation does not capture correctly the structure
of the flow in a neighbourhood of the body  where $\mathcal{D}\geq c_{\infty}^{2}R_{p}/(GM)$.
Indeed, the structure of the flow in the non-linear region depends on whether the perturber 
is modeled as a perfect accretor or, on the opposite case, it is modeled as a non-divergent core 
gravitational potential. 

\begin{figure}
\epsfig{file=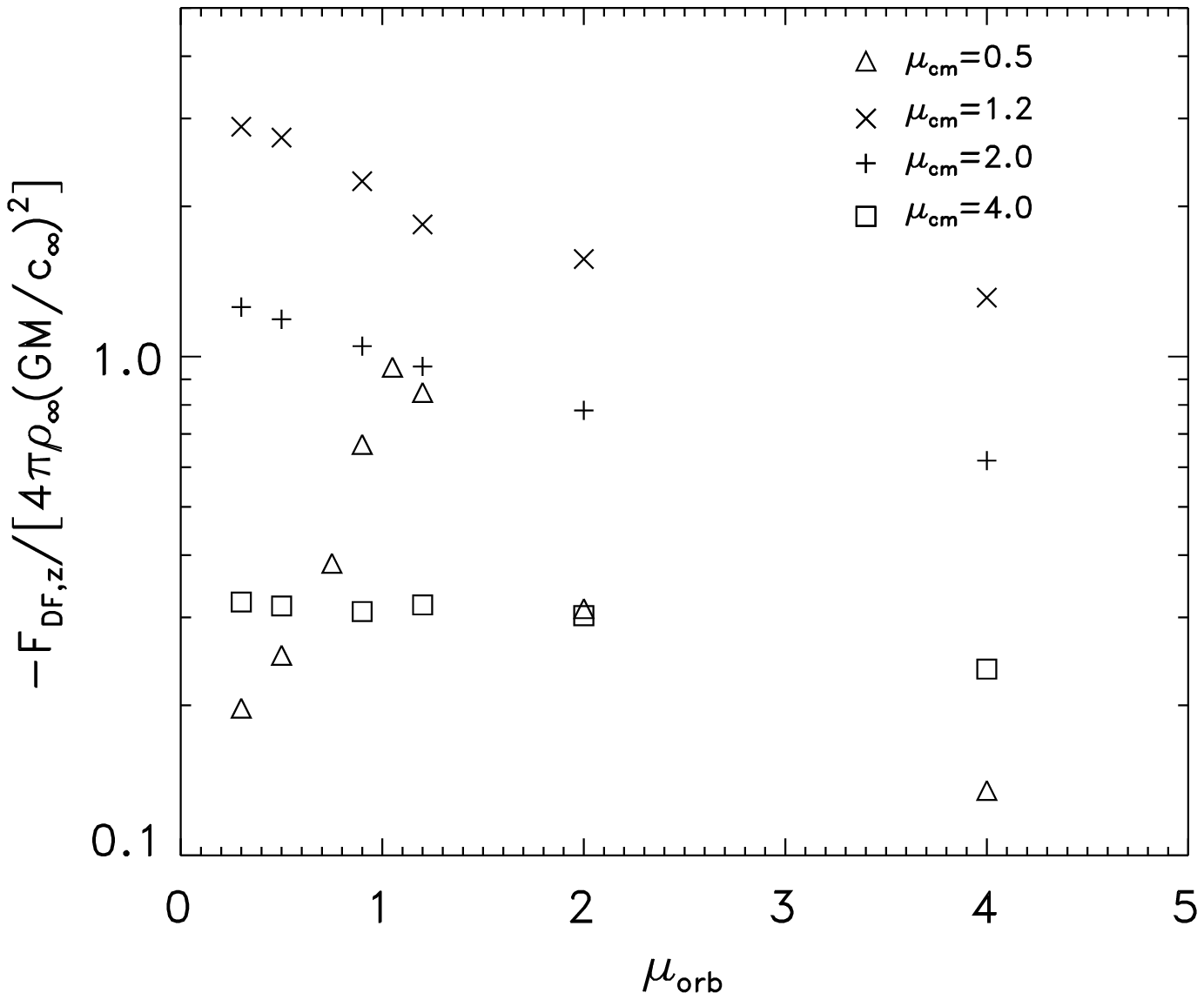,angle=0,width=15.5cm}
  \caption{Vertical component (i.e. along the $z$-direction)
of the dimensionless dynamical friction force for a gravitational
object in the face-on case, as a function of $\mu_{\rm orb}$.  }
\label{fig:fz_vs_orb_face}
\vskip 0.75cm
\end{figure}

\section{The wake induced by one component }
\label{sec:wake}
We specialize in the case where the perturber is introduced at $t=0$. This corresponds to
take $\hcur(t)=\Theta(t)$, where $\Theta(t)$ is the Heaviside step function, so that at $t<0$ the
medium is unperturbed.  As a result, the
summation in Equation (\ref{eq:D_general}) is only over positive roots (that is, only roots
$\phi_{j}>0$ contribute in Equation \ref{eq:D_general}); in fact, $\hcur$ defines
the region that sound waves launched at $t=0$ have time to reach.

\subsection{Face-on case}
\label{sec:wakefaceon}

The temporal evolution of the wake created
by a perturber in a pure circular orbit, i. e. $\mu_{\rm cm}=0$, was described in full detail in 
Kim \& Kim (2007). Due to the orbit curvature, 
the wake tail bends behind the perturber, creating
spiral waves. As illustrated by Kim \& Kim (2007), the thickness of the spiral waves and their
level of winding depend on the orbital Mach number. Here we consider the case where $\VV_{\rm cm}$
is nonzero.

As said in Section \ref{sec:orbit}, when $i=0$, the orbital plane is perpendicular to $\VV_{\rm cm}$ and
the body describes a helical trajectory. It is interesting to compare the morphologies of the wakes created
by  perturbers in helical motion with those created by perturbers in circular orbit.
Figure \ref{fig:mu_subsonic} shows the dimensionless density $\mathcal{D}$ at $\tilde{t}=9$
(remind that the time unit is $R_{p}/c_{\infty}$),
when $\mu_{\rm orb}=0.5$ for two values of $\mu_{\rm cm}$ ($0$ and $0.5$).
In both cases, the structure of the density wake presents a 
comma-like shape at the plane of the orbit, i.e. at $\tilde{z}=0$ when $\mu_{\rm cm}=0$ and
at $\tilde{z}=4.5$ when $\mu_{\rm cm}=0.5$. Since the perturbers move subsonically 
($\mu_{\rm orb}^{2}+\mu_{\rm cm}^{2}<1$) in both cases,
they always lie in the interior of the sonic sphere. Close enough to the
perturber, at distances $\lesssim R_{p}$, a spheroidal envelope is formed around the perturber. At larger
distances from the perturber, the disturbance forms a trailing one-armed spiral wave. This one-armed
disturbance ends at the sonic sphere.  
The density maps in the $(x,y)$-plane that contains the perturber,
look like qualitatively similar in both cases. In these maps,
the perturbations are confined to a circle of radius $9R_{p}$
in the plane $\tilde{z}=0$, and to a circle of radius $7.8 R_{p}$ in the plane $\tilde{z}=4.5$.
These circles are obtained as the intersection of the sonic
sphere, which is centered at the initial position of
the perturber and has a radius of $9R_{p}$, with the planes $\tilde{z}=0$ and $\tilde{z}=4.5$, respectively.
Interestingly, the density maps along vertical cuts are remarkably different.
For $\mu_{\rm cm}=0$, the density perturbation is symmetric with respect to the plane $z=0$. 
This symmetry is broken when the perturber moves in the vertical direction; an elongated overdense
structure can be observed in the cut along a vertical plane when $\mu_{\rm cm}=0.5$.

The morphology of the wake changes dramatically for the perturber moves
supersonically in the vertical direction. 
In the extreme case where $\mu_{\rm cm}>1$ and $\mu_{\rm cm}\gg \mu_{\rm orb}$,  
a Mach cone is formed behind the perturber.
However, because of the orbital motion, the axis of the Mach cone is not a straight line,
as occurs in the rectilinear case, 
but follows a helicoidal curve. This occurs, for instance, for $\mu_{\rm orb}=0.5$ and 
$\mu_{\rm cm}=2$ (not shown).

Now consider the case where the orbital motion is supersonic $\mu_{\rm orb}>1$.
For $\mu_{\rm cm}=0$, the gravitational disturbance leads to the formation of an
one-arm trailing tail
that wraps around the perturber (see Kim \& Kim 2007). The radial separation between
two consecutive spiral crests is $2\pi R_{p}\mu_{\rm orb}^{-1}$.
Figures \ref{fig:muorb2_horizontal} and \ref{fig:muorb2_vertical} display the density disturbance
 for $\mu_{\rm orb}=2$ and different $\mu_{\rm cm}$-values, 
along horizontal and vertical planes, respectively. 
When the perturber moves in the vertical direction, it is easy to identify the corresponding
spiral wave fronts, more clearly in cuts along $z=$const planes, as long as $\mu_{\rm cm}<1$.
For supersonic motions in the vertical direction ($\mu_{\rm cm}>1$),
the morphology of the wake changes remarkably (see the fourth panel in Figure \ref{fig:muorb2_vertical}).

Figure \ref{fig:muorb2_vertical} also shows the transversal structure of the spiral waves, which appear as long arcs 
(for $\mu_{\rm cm}<1$). For $\mu_{\rm cm}\rightarrow 1$,
the spiral wave fronts are bunched up closer together at $\mu_{\rm cm}\tilde{t}<\tilde{z}<\tilde{t}$
(i.e. at $V_{\rm cm}t<z<c_{\infty}t$ in physical units), especially
for $\mu_{\rm cm}=0.9$, because of the vertical motion of the perturber. 
Since we have selected vertical cuts that 
pass through the position of the perturber, we can see the Mach cone at the rear of the perturber,
at distances $\sim R_{p}$ from it. Indeed, the Mach cone is formed because the perturber
moves supersonically relative to the background. 

When $\mu_{\rm cm}=\mu_{\rm orb}=2$, the vertical distribution of the wake is very complex, with many
overdense substructures. However, the overall density perturbation is confined within 
the large-scale rear Mach cone, defined by $R<R_{p}-(z-z_{p})/(\mu_{\rm cm}^{2}-1)^{1/2}$,
where $R^{2}=x^{2}+y^{2}$ is the cylindrical distance (see Appendix \ref{sec:app_location}).
In fact, the perturber launches sound waves to the medium
that necessarily must lie within the large-scale Mach cone to preserve the causality condition.

\subsection{Edge-on case}
The total Mach number $(\dot{x}_{p}^{2}+\dot{y}_{p}^{2}+\dot{z}_{p}^{2})^{1/2}/c_{\infty}$ 
is constant along the orbit of the perturber only when $i=0$. 
Otherwise, the total Mach number varies with the perturber's angular distance $\phi_{p}$
measured in the orbital plane, defined as $\phi_{p}\equiv \Omega t$ (see Eqs. \ref{eq:xp}-\ref{eq:zp}).
In this Section, we will focus on the edge-on orbit, $i=\pi/2$.
In such a case, $y_{p}=0$ at any time. 
The body moves in the plane $y=0$ on an epicycle, which in turn moves along the $z$-axis.
The maximum
total Mach number is $\mu_{\rm cm}+\mu_{\rm orb}$ and occurs when $\phi_{p}=2n\pi $ (with
$n$ an integer).
The minimum total Mach number is $|\mu_{\rm cm}-\mu_{\rm orb}|$ at $\phi_{p}=(2n+1)\pi$.

Figure \ref{fig:D_edgeon} shows the map of the density disturbance in the plane $y=0$ for different
combinations of $\mu_{\rm cm}$ and $\mu_{\rm orb}$.  We see that when
$\mu_{\rm cm}+\mu_{\rm orb}<1$, that is, when the motion of the perturber is always subsonic,
a comma-type wave, similar to that found in the face-on case with 
$\mu_{\rm cm}^{2}+\mu_{\rm orb}^{2}<1$, is formed.
It is remarkable that when $\mu_{\rm cm}=0.5$ and $\mu_{\rm orb}=2$,
a spiral wave also emerges, as occurs in the face-on orbit, but now the thickness 
of the spiral wave clearly varies with $z$, being thicker at $z> z_{p}$ and thinner
at $z<z_{p}$. This can be interpreted as follows. The spiral sound 
waves moves radially outwards at a velocity $c_{\infty}$. Thus, the relative
velocity between the perturber and the spiral waves in the radial direction
varies with the azimuthal angle $\phi$, from $(1-\mu_{\rm cm})c_{\infty}$
at $\phi_{p}=\pi/2$, to $(1+\mu_{\rm cm})c_{\infty}$, at $\phi_{p}=-\pi/2$.
This asymmetry is responsible for the asymmetric thickness of the spiral waves
found when $\mu_{\rm orb}>1$ and $\mu_{\rm cm}<1$.

When $\mu_{\rm cm}=2$ and $\mu_{\rm orb}=0.5$,  the velocity of the perturber
along the $z$-direction is always supersonic and, hence, the perturber is always at the
apex of the deformed Mach cone.
 The overall structure of the wake resembles the wake of a supersonic body in the
straight-line trajectory case but now the Mach cone is deformed because of the epicyclic motion and
also because the total Mach number is not constant but varies between $1.5$ and $2.5$ along one 
epicycle.

The structure of the wake is very complex when $\mu_{\rm cm}$ and $\mu_{\rm orb}$ are
both larger than $1$ and comparable in magnitude. As an example, we show the case for
$\mu_{\rm cm}=\mu_{\rm orb}=1.2$. The total Mach number oscillates
between $0$ and $2.4$. Density disturbances lag the perturber when
it moves at the highest total Mach number ($2.4$ in this case), which occurs at $\phi_{p}=2n\pi$. 
However, the perturber is swallowed by its own wake when it moves very subsonically 
at $\phi_{p}\approx (2n+1)\pi$, creating a wake with a very irregular morphology.

\begin{figure}
\includegraphics[width=130mm,height=140mm]{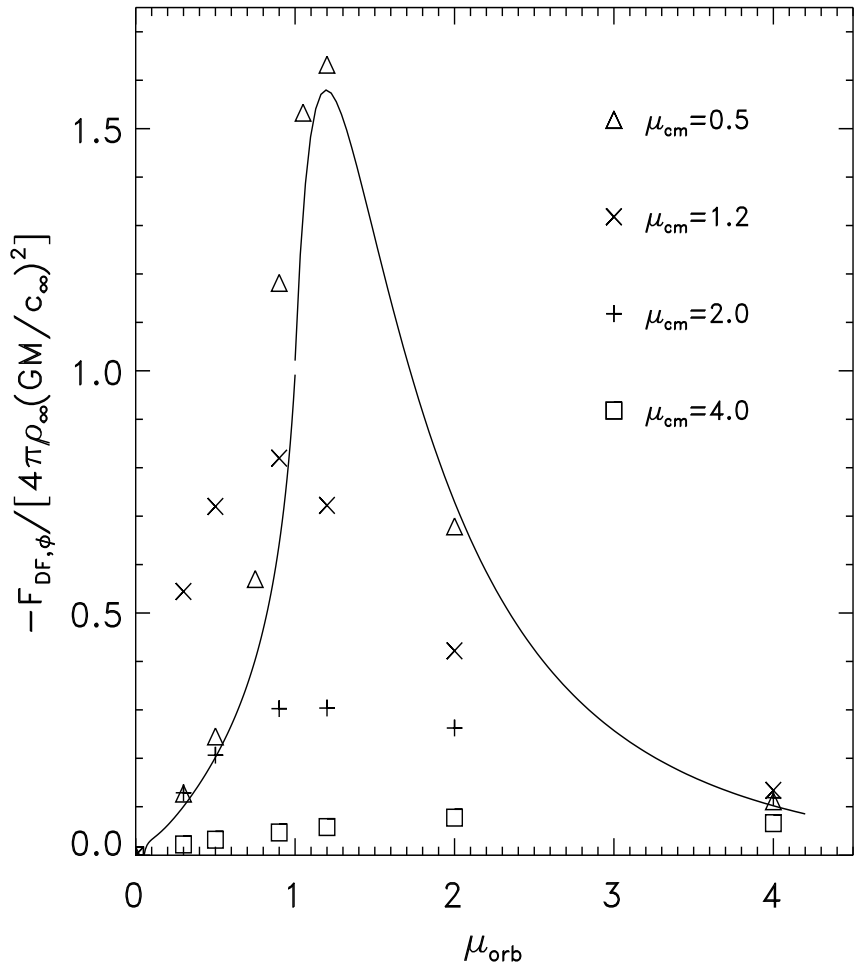}
  \caption{Azimuthal dynamical friction force, in dimensionless units, against $\mu_{\rm orb}$
in the face-on case. The force was computed at $\tilde{t}=9$. At those times, the azimuthal
component has practically reached the value of saturation.
The solid line corresponds to the analytical fit derived by Kim \& Kim (2007)
for a pure circular orbit ($\mu_{\rm cm}=0$).}
\vskip 0.75cm
\label{fig:fphi_vs_orb_face}
\end{figure}

\section{The dynamical friction force}
\label{sec:drag}
When a gravitational body moves through a gaseous medium, it experiences two forces, the aerodynamic
force due to accretion, and the dynamical friction force, which arises from the gravitational
attraction between the perturber and its wake.

In the case of a binary system, each component produces its own wake. Let the components
of the binary have masses $M$ and $fM$, where $f\leq 1$. In linear theory, the density
disturbance in the wake is $\alpha=\alpha_{1}+\alpha_{2}$, where $\alpha_{1}$
and $\alpha_{2}$ are the wakes induced by the perturbers with mass $M$ and $fM$, respectively.
Once the density wake $\alpha(\rr,t)$ is known, the dynamical friction force exerted on
the perturber of mass $M$ can be computed as:
\begin{equation}
\FF_{DF}^{(1)}=GM \rho_{\infty} \int  \frac{\alpha(\rr,t)(\rr-\rr_{p,1})}
{|\rr-\rr_{p,1}|^{3}}d^{3}\rr,
\label{eq:integral_force}
\end{equation}
where $\rr_{p,1}$ is the position vector of the perturber of mass $M$. 
Similarly, we can obtain $\FF_{DF}^{(2)}$,
the force acting on the mass $fM$.

Since in linear theory $\alpha$ diverges when $\rr \rightarrow \rr_{p,1}$, it is customary to 
introduce a minimum cut-off radius $r_{\rm min}$ to regularize the integral (\ref{eq:integral_force}).
Although there is some ambiguity in the definition of $r_{\rm min}$, in practice, it is 
taken as the characteristic distance from the body where the linear
approximation fails (that is, $r_{\rm min}\approx r_{\rm nl}$). For instance,  
it is well-documented that for a compact object in 
supersonic rectilinear orbit, the linear formula for the drag force reproduces the
drag force once the value provided for $r_{\rm min}$ is 
the accretion radius (e.g., Cant\'o et al. 2011; Bernal \& S\'anchez-Salcedo 2013).

\begin{figure}
\epsfig{file=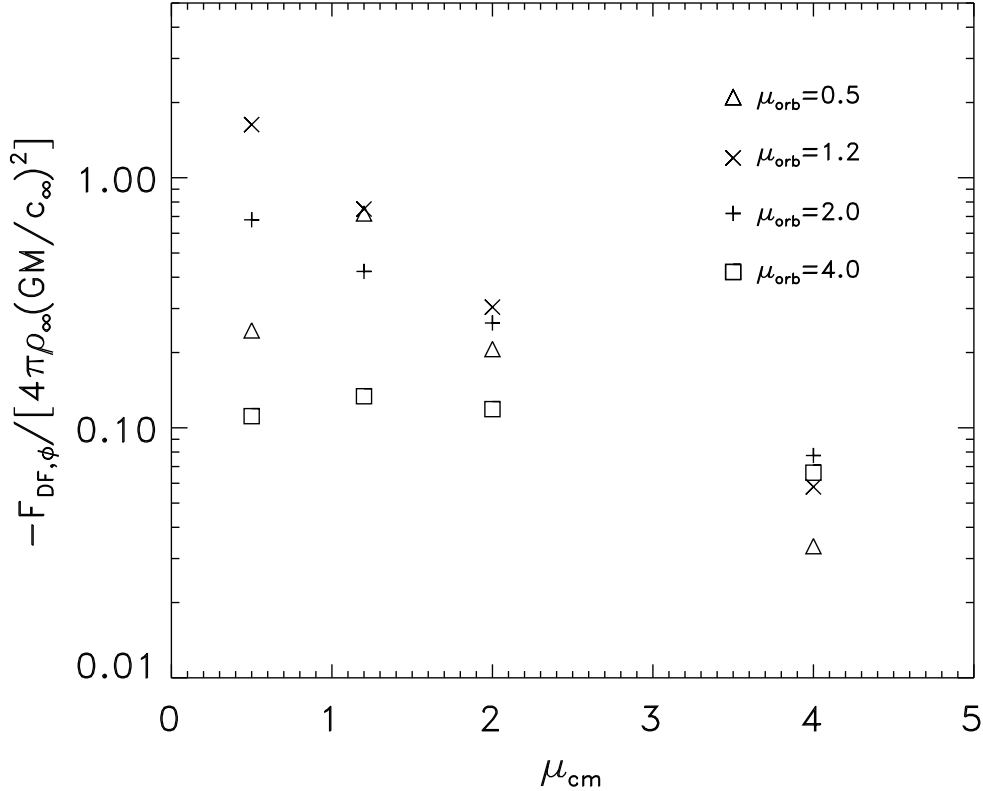,angle=0,width=14.5cm}
  \caption{Azimuthal component of the drag force, as a function of $\mu_{\rm cm}$, for a
body in face-on motion.  It was computed at $\tilde{t}=9$.}
\label{fig:fphi_vs_com_face}
\vskip 0.75cm
\end{figure}

\subsection{Components of the force}
\label{sec:components}
The binary is subject to a net force $\FF_{DF}^{(1)}+\FF_{DF}^{(2)}$, hence
\begin{equation}
(1+f)M\frac{d\VV_{\rm cm}}{dt}=\FF_{DF}^{(1)}+\FF_{DF}^{(2)}.
\end{equation}
One expects that the net force will tend to decelerate the binary center of mass, which we have
assumed that moves in the $z$-direction.  
For $i=0$ and $f=1$, $\FF_{DF}^{(1)}+\FF_{DF}^{(2)}$ points in the $z$-direction
because of the symmetry of the problem.

As we have seen in Section \ref{sec:wake}, binary systems induce spiral waves in the ambient
medium, which are suitable to produce a braking torque on the binary. 
The angular momentum of the binary about the center of mass, $\LL$, evolves according to
\begin{equation}
\frac{d\LL}{dt}=\rr_{p,1}\times \FF_{DF}^{(1)}+\rr_{p,2}\times \FF_{DF}^{(2)}.
\end{equation}
Torques perpendicular to $\LL$ cause the binary's orbital plane to precess.
Here we are interested in the change of $L^{2}$ due to the gravitational interaction
with the surrounding gas. 
If the binary gains angular momentum it widens, whereas it shrinks when it loses angular momentum.
The variation of $L$ with time is given by
\begin{equation}
\frac{dL^{2}}{dt}=
2[\FF_{DF}^{(1)}\cdot(\LL\times \rr_{p,1})+\FF_{DF}^{(2)}\cdot(\LL\times \rr_{p,2})].
\end{equation}
If the two components of the binary are on circular orbits about their common center of 
mass\footnote{As usual, we assume that the change of the orbital parameters of the binary
due to the drag force and internal torques on the binary
occurs on a timescale much longer than any other relevant timescale.
So, it is a good approximation to consider unperturbed orbits.},
$\LL$ is perpendicular
to both $\rr_{p,1}$ and $\rr_{p,2}$.  Hence, we can write $\LL\times \rr_{p,1}=R_{p,1}L\hat{\ee}_{\phi,1}$,
where $\hat{\ee}_{\phi,1}$ is the vector perpendicular to both $\LL$ and $\rr_{p,1}$.
We arrive at
\begin{equation}
\frac{dL}{dt}=R_{p,1}F_{DF,\phi}^{(1)}+R_{p,2}F_{DF,\phi}^{(2)}.
\end{equation}

In order to easy the comparison with previous studies that deal with a single perturber, 
we provide first $\FF_{DF}$, defined as the force exerted on the perturber of mass $M$ 
by its own induced wake ${\mathcal{D}}_{1}$:
\begin{equation}
\FF_{DF}\equiv \frac{(GM)^{2}\rho_{\infty}}{c_{\infty}^{2}}  \int  \frac{{\mathcal{D}}_{1}(\rr,t)(\tilde{\rr}-\hat{\rr}_{p,1})}
{|\tilde{\rr}-\hat{\rr}_{p,1}|^{3}}d^{3}\tilde{\rr}.
\label{eq:force_own}
\end{equation}
To simplify the notation, the subscript $1$ to denote perturber's orbital radius will be
dropped out: $R_{p,1}\rightarrow R_{p}$. 

In order to evaluate the integral in Eq. (\ref{eq:force_own}), the computational domain was
divided in several blocks, along the wake, with different degree of refinement; the block containing
the perturber has the highest resolution (about $200$-$400$ zones per $R_{p}$). The resolution of
the blocks was degraded depending on the distance to the perturber. For the most demanding calculations,
those having large Mach numbers or Mach numbers close to unity, $8$ blocks were used.
For supersonic perturbers, we introduced a small softening of $\sim 2$ pixels at discontinuities
where ${\mathcal{D}}$ diverges, in order to suppress numerical noise.
Note that linear theory predicts those kind of density discontinuities  
even in the linear trajectory case (e.g., Ostriker 1999).

\subsection{Drag force induced by its own wake: face-on orbits}
In this Section, we consider one component of the binary and report the drag force due
to its own wake $\FF_{DF}$ in the face-on case ($i=0$). $\FF_{DF}$ will depend on three
parameters: $\mu_{\rm orb}, \mu_{\rm cm}$ and the adopted value for $r_{\rm min}$.
As already discussed, $r_{\rm min}$ is connected with the different gravitational
spheres of influence defined in Section \ref{sec:scales}. For practical purposes,
the $r_{\rm min}$ value
must be calculated in each particular situation. For a circular binary with masses $M$ and $fM$
in a medium with sound speed $c_{\infty}$, the cut-off radius is entirely determined by $f$, $c_{\infty}$,
$\mu_{\rm orb}$ and $\mu_{\rm cm}$.

In order to isolate the effect of each three parameter $\mu_{\rm orb}$, $\mu_{\rm cm}$ and
$r_{\rm min}$, we first study the dependence of the drag force on the Mach numbers
using a fixed value of $r_{\rm min}$ (Section \ref{sec:muDependence}).  This is customary
in studies of dynamical friction and will help us to obtain a more complete understanding on
how the presence of epicyclic motions modify the nature of the force.
We will take $r_{\rm min}=0.1R_{p}$ to
facilitate comparison with Kim \& Kim (2007). 
The dependence of the drag force on $r_{\rm min}$ is discussed in Section \ref{sec:rmin}.

\subsubsection{Drag forces: Dependence on $\mu_{\rm cm}$ and $\mu_{\rm orb}$}
\label{sec:muDependence}

Figure \ref{fig:force_vs_time} shows the components
of $\FF_{DF}$, as functions of time, for $\mu_{\rm cm}=2$ and $\mu_{\rm orb}=0.9$ (remind that,
unless otherwise state, we use $r_{\rm min}=0.1R_{p}$). 
Remind that $F_{DF,\phi}$ is the azimuthal component and $F_{DF,R}$ the radial
component in the frame of the binary, with its center of mass situated at the origin.
We see
that $F_{DF,\phi}$ and $F_{DF, R}$ both converge to their respective steady-state values, but
$F_{DF,z}$ increases in time logarithmically.
Kim \& Kim (2007) already found that for a perturber on a pure circular orbit ($\mu_{\rm cm}=0$),
the drag components $F_{DF,\phi}$ and $F_{DF,R}$ converge to their respective steady-state
values within a timescale $\sim 2R_{p}/c_{\infty}$ for subsonic perturbers or within 
$\sim 2\pi R_{p}/V_{\rm orb}$ for supersonic perturbers. 
This means that the temporal behaviour of $F_{DF,\phi}$
and $F_{DF, R}$ is similar to the case of pure circular orbit, whereas the temporal behaviour
of $F_{DF, z}$ is similar to the case of rectilinear orbit. The reason is clear: in the 
$z$-direction, the Mach cone trailing the perturber is continuously growing along
the $z$-direction because $\mu_{\rm cm}>1$. In the $\phi$-direction, the curvature of
the orbit produces the saturation of $F_{DF,\phi}$ and $F_{DF,R}$. When $\mu_{\rm cm}<1$,
the three components of $\FF_{DF}$ asymptotically approach constant values with time.

Figure \ref{fig:fz_vs_com_face} presents the vertical drag component, $F_{DF,z}$,
 at $\tilde{t}=9$ as a function of $\mu_{\rm cm}$ for
different values of $\mu_{\rm orb}$.
In order to quantify the effect of the orbital motion on $F_{DF,z}$, we have also plotted $F_{DF,z}$ 
for a body with $\mu_{\rm orb}=0$:
\begin{equation}
F_{DF,z}=\frac{4\pi \rho_{\infty}(GM)^{2}}{c_{\infty}^{2}}
\frac{1}{\mu_{\rm cm}^{2}}\left[\frac{1}{2}\ln\left(\frac{1+\mu_{\rm cm}}{1-\mu_{\rm cm}}\right)-
\mu_{\rm cm}\right],
\end{equation}
if $\mu_{\rm cm}=0$, $\mu_{\rm cm}<1$ and $t>r_{\rm inf}/[c_{\infty}(1-\mu_{\rm cm})]$, and 
\begin{equation}
F_{DF,z}=\frac{4\pi \rho_{\infty}(GM)^{2}}{c_{\infty}^{2}}
\frac{1}{\mu_{\rm cm}^{2}}\left[
\frac{1}{2}\ln (1-\mu_{\rm cm}^{-2})+\ln\left(\frac{\mu_{\rm cm}c_{\infty}t}{r_{\rm inf}}\right)\right],
\end{equation}
if $\mu_{\rm cm}=0$, $\mu_{\rm cm}>1$ and $t>r_{\rm inf}/[c_{\infty}(\mu_{\rm cm}-1)]$
(Ostriker 1999). Here $r_{\rm inf}$ is the minimum cut-off radius for a particle in a straight-line
orbit. This implies that when $\mu_{\rm orb}=0$, $r_{\rm inf}$ must be equal to $r_{\rm min}$. 
In fact, Figure \ref{fig:fz_vs_com_face} illustrates this situation; 
for $\mu_{\rm orb}<0.5$, Ostriker's formula with $r_{\rm inf}=r_{\rm min}$
provides reasonably good estimates of the vertical drag force.

Figures \ref{fig:fz_vs_com_face} and \ref{fig:fz_vs_orb_face} show that, at a fixed value of $\mu_{\rm cm}$ greater than $1$, 
$F_{DF,z}$ decreases when $\mu_{\rm orb}$ increases. The physical reason is that the orbital motion
around the center of mass,
induces a loss of gravitational focusing in the $z$-direction. If we wish to continue using Ostriker's
formula, we must use a larger effective $r_{\rm inf}$. For instance, Ostriker's formula provides
the correct value of $F_{DF,z}$, for $\mu_{\rm orb}=0.9$ and $\mu_{\rm cm}>1$, when
$r_{\rm inf}=2.25r_{\rm min}$ is used. For $\mu_{\rm orb}=1.2$, one requires 
$r_{\rm inf}=3.36r_{\rm min}$.

In Figure \ref{fig:fz_vs_orb_face}, we see that at a fixed $\mu_{\rm cm}=0.5$, the steady-state
vertical component $F_{DF,z}$, as a function of $\mu_{\rm orb}$, 
has its peak at around $\mu_{\rm orb}=1.2$, being its magnitude almost $4$ times
larger than it is for $\mu_{\rm orb}=0.3$ or $\mu_{\rm orb}=4$. Beyond $\mu_{\rm orb}=1.2$,
$F_{DF,z}$ decreases rapidly with $\mu_{\rm orb}$.

The azimuthal component of the force, $F_{DF,\phi}$, which is responsible for the loss
of orbital angular momentum, is shown in Figures \ref{fig:fphi_vs_orb_face}
and \ref{fig:fphi_vs_com_face}. 
For $\mu_{\rm cm}=0$, our values
of $F_{DF,\phi}$ match those derived by Kim \& Kim (2007).
At $\mu_{\rm cm}=0.5$, $F_{DF,\phi}$ is slightly shifted as compared
to the values for $\mu_{\rm cm}=0$. Therefore, the analytical formula in
Kim \& Kim (2007) provides good
estimates of $F_{DF,\phi}$ as long as $\mu_{\rm cm}<0.5$. 
In general, Kim \& Kim (2007) formula is satisfactory
when $\mu_{\rm orb}\gg \mu_{\rm cm}$.

Figure \ref{fig:fphi_vs_orb_face} shows the 
functional relationship between $F_{DF,\phi}$ and $\mu_{\rm orb}$ for different
values of $\mu_{\rm cm}$. We see that,
in the range $0<\mu_{\rm orb}<2.5$, the $F_{DF,\phi}$-$\mu_{\rm orb}$ relationship 
derived at $\mu_{\rm cm}=1.2$ is
different from the relationship found for $\mu_{\rm cm}=0$.
The maximum of $F_{DF,\phi}$ for $\mu_{\rm cm}=1.2$ occurs
at orbital mach numbers around $0.9$ and its value is a factor of $2$ smaller than the 
corresponding peak value of $F_{DF,\phi}$ vs. $\mu_{\rm orb}$ at $\mu_{\rm cm}=0$.

For subsonic orbital Mach numbers ($\mu_{\rm orb}<1$), $F_{DF,\phi}$ is significantly
larger for $\mu_{\rm cm}=1.2$
than it is for $\mu_{\rm cm}=0$ (see Figure \ref{fig:fphi_vs_orb_face}).
In particular, $F_{DF,\phi}$  for $\mu_{\rm orb}=0.5$ and $\mu_{\rm cm}=1.2$
is a factor of $3.5$ larger than it is for $\mu_{\rm orb}=0.5$ and $\mu_{\rm cm}=0$.
The reason is that the high front-back symmetry of the wake near the body 
(say, at distances $\leq R_{p}$), when
$\mu_{\rm cm}=0$ and $\mu_{\rm orb}=0.5$, is broken when the perturber moves supersonically 
(as $\mu_{\rm cm}=1.2$). Therefore, it is misleading to think that the presence
of translational motions
always reduces $F_{DF,\phi}$. For $\mu_{\rm orb}<1$, $F_{DF,\phi}$ with
$\mu_{\rm cm}=2$ is very similar to $F_{DF,\phi}$ for $\mu_{\rm cm}=0$.

When the orbital and the vertical Mach numbers
are similar and supersonic,  $\mu_{\rm cm}\simeq \mu_{\rm orb}>1$, the azimuthal component 
of the drag force is
reduced by a factor of $2$ as compared to the case when $\mu_{\rm cm}=0$.
As expected, the largest suppression in $F_{DF,\phi}$ occurs at the largest $\mu_{\rm cm}$.
For $\mu_{\rm orb}=1.2$ and $\mu_{\rm cm}=4$, $F_{DF,\phi}$ is a factor of $30$ smaller than it
is for $\mu_{\rm orb}=1.2$ and $\mu_{\rm cm}=0$.

Figure \ref{fig:fphi_vs_com_face} shows $F_{DF,\phi}$ as a
function of $\mu_{\rm cm}$ for different values of $\mu_{\rm orb}$. When $\mu_{\rm orb}=4$,
$F_{DF,\phi}$ only varies about $30\%$ in the range $0<\mu_{\rm cm}<4$. However, when
$\mu_{\rm orb}=1.2-2$, $F_{DF,\phi}$ decreases monotonically with $\mu_{\rm cm}$  by
a factor $\geq 10$ in the range $0<\mu_{\rm cm}<4$.

\begin{figure}
\epsfig{file=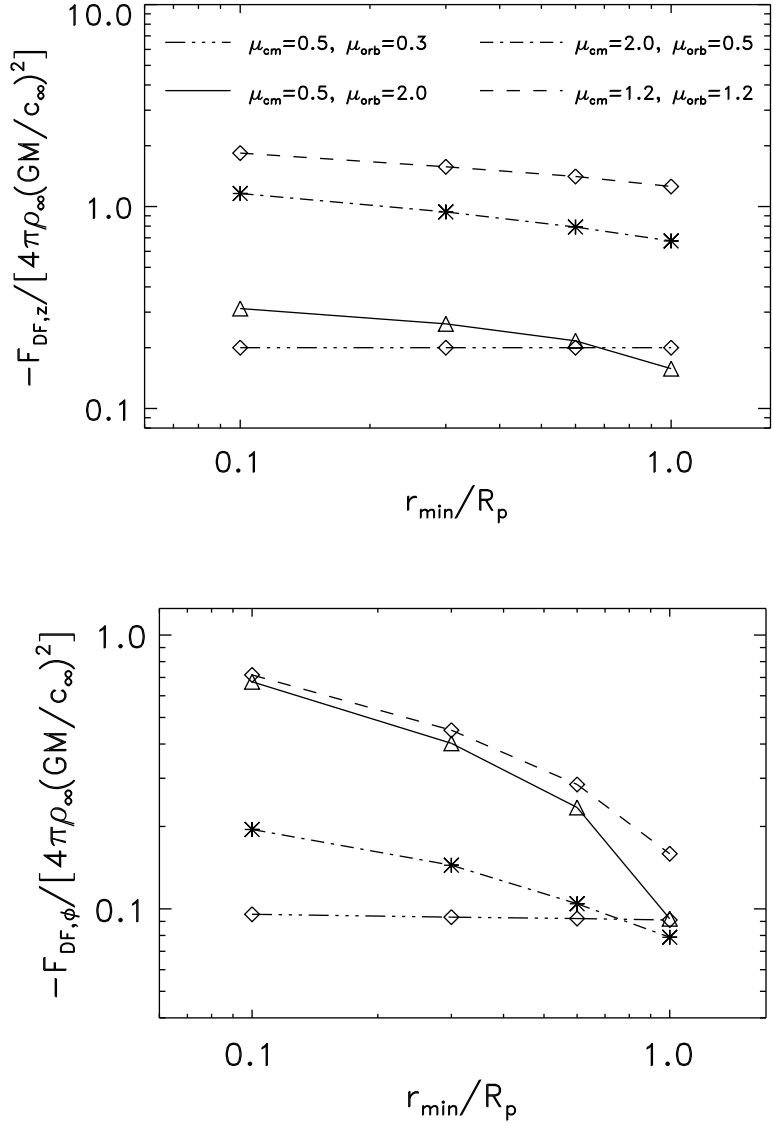,angle=0,width=11.5cm}
  \caption{Dependence of the $z$-component (upper panel) and azimuthal component (lower panel)
of the drag force on the adopted value of $r_{\rm min}$, for four different combinations
of $\mu_{\rm cm}$ and $\mu_{\rm orb}$. The forces were calculated at $\tilde{t}=9$. }
\label{fig:rmin}
\vskip 0.75cm
\end{figure}

\subsubsection{Drag forces: Dependence on $r_{\rm min}$}
\label{sec:rmin}

Figure \ref{fig:rmin} shows the vertical and azimuthal components
of the drag force for different values of $r_{\rm min}$, and for some
combinations of $\mu_{\rm cm}$ and $\mu_{\rm orb}$. When both motions
are subsonic, the drag force is not sensitive to the value of
$r_{\rm min}$ as it varies from $0.1R_{p}$ to $1 R_{p}$. Physically, this means
that the wake within $R_{p}$ does not contribute to the drag because
of its back-front symmetry.

It is useful to define $\Delta F_{z}$ as the contribution to the (dimensionless) drag
force by the portion of the wake lying between a distance $0.1R_{p}$ and a distance
$R_{p}$ from the perturber. 
For a supersonic perturber moving in a straight-line with Mach number $\mu$,
$\Delta F_{z}=\mu^{-2} \ln 10$. This implies that
$\Delta F_{z}=1.6$ for $\mu=1.2$, and $\Delta F_{z}=0.57$ at $\mu=2$. 
In the case of helical motion with
$\mu_{\rm cm}=2$ and $\mu_{\rm orb}=0.5$, we found that 
$\Delta F_{z}=0.48$ (see Fig. \ref{fig:rmin}), which
is slightly smaller than for the rectilinear orbit. However, for $\mu_{\rm cm}=1.2$ and 
$\mu_{\rm orb}=1.2$, we found that $\Delta F_{z}=0.55$ if the motion is helical.
This value is much less than the value for rectilinear orbit ($1.6$) because the gravitational
focusing of mass is reduced at distances $\sim R_{p}$ from the perturber due to the
orbital (epicyclic) motion.

\begin{figure*}
\begin{centering}
\epsfig{file=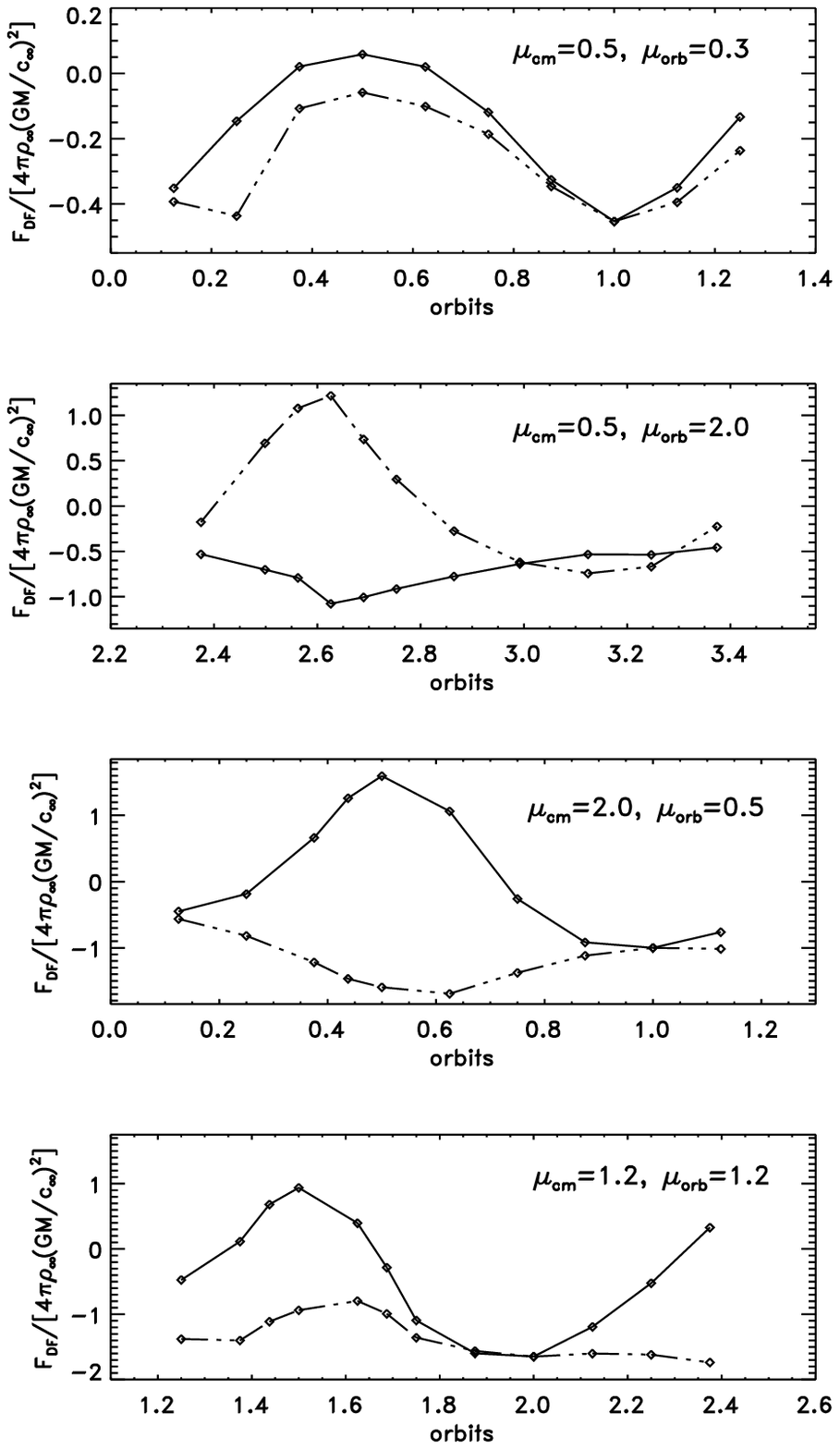,angle=0,width=11cm,height=19.5cm}
\caption{Azimuthal component  $F_{DF,\phi}$ (solid lines) and
$z$-component (dash-dotted line) of the drag force for an edge-on perturber
($i=90^{\circ}$) along approximately one orbit, for different combinations of $\mu_{\rm cm}$ and
$\mu_{\rm orb}$. }
\label{fig:force_edge}
\end{centering}
\vskip 0.75cm
\end{figure*}

The dependence of $F_{DF,\phi}$ on $r_{\rm min}$ is shown in the lower panel of Fig. \ref{fig:rmin}.
We see that the slope of these curves increases with $\mu_{\rm orb}$.
Still, even when the orbital motion is subsonic,  $F_{DF,\phi}$ is sensitive to $r_{\rm min}$
if the translational motion is large enough. For instance, for $\mu_{\rm orb}=0.5$
and $\mu_{\rm cm}=2$, $F_{DF,\phi}$ decreases a factor of $2.4$ if $r_{\rm min}=R_{p}$,
instead of $0.1R_{p}$, is used. It is noteworthy that 
for $\mu_{\rm orb}=2$ (and $\mu_{\rm cm}=0.5$), $F_{DF,\phi}$ decreases
by a factor of $\sim 7$ on the interval of $r_{\rm min}$ under consideration. This signifies
that most of the azimuthal component of the drag force is caused by the gravitational
attraction between the perturber and gas within a volume of radius $\sim R_{p}$ surrounding the perturber.
Therefore, $F_{DF,\phi}$ is very sensitive to the adopted value of $r_{\rm min}$. 

The strong suppression of $F_{DF,\phi}$ with $r_{\rm min}$ at $\mu_{\rm orb}\gtrsim 2$
and $\mu_{\rm cm}\lesssim 1$ has profound implications when dealing with binary systems. 
To illustrate this issue, consider an equal-mass binary. According to Figure \ref{fig:parameter_space},
$r_{\rm min}\gtrsim 10 R_{p}$ at the abovementioned Mach numbers. Unfortunately, we were
unable to obtain accurate inferences of $F_{DF,\phi}$ when using $r_{\rm min}>10R_{p}$ because 
the result is severely polluted by numerical noise. Nevertheless, a rough extrapolation of the solid
line in Figure \ref{fig:rmin} at $r_{\rm min}=10 R_{p}$ indicates that the contribution of the ``linear'' 
part of the wake to the azimuthal drag force, is very small. Under these circumstances, it is possible 
that the contribution of the nonlinear part of the wake dominates the value of $F_{DF,\phi}$.
This possibility can only be tested by using fully hydrodynamical simulations.

\subsection{Drag force induced by its own wake: edge-on orbits}
Consider now the extreme case $i=\pi/2$, where the orbital plane lies in the $x-z$ plane and
the center of mass moves along the $z$-direction. By symmetry, $F_{DF,y}=0$. Thus, the drag
force lies within the plane $y=0$. In this case, the components $F_{DF,z}$ and $F_{DF,\phi}$ 
will depend on the perturber's azimuthal angle $\phi_{p}$ (remind that $\phi_{p}\equiv \Omega t$).
Figure \ref{fig:force_edge} shows $F_{DF,z}$ and $F_{DF,\phi}$ along one orbital time $t_{\rm orb}$, which
is $2\pi/\Omega=2\pi R_{p}/(c_{\infty}\mu_{\rm orb})$. 
We have used $r_{\rm min}=0.1R_{p}$. The reference time $\tilde{t}=9$,
corresponds to $0.43$ orbits when $\mu_{\rm orb}=0.3$, to $2.86$ orbits when $\mu_{\rm orb}=2$,
to $0.72$ orbits when $\mu_{\rm orb}=0.5$ and to $1.72$ orbits when $\mu_{\rm orb}=1.2$.

When $\mu_{\rm cm}=0.5$ and $\mu_{\rm orb}=0.3$, the two components $F_{DF,z}$ and
$F_{DF,\phi}$, vary approximately in phase. They take their maximum absolute value at $\phi_{p}=2\pi$,
i.e.~when an orbit is completed and the instantaneous total Mach number is $0.8$, whereas
the minimum absolute values occur at $\phi_{p}=\pi$,  when the total Mach number is $0.2$.
The mean values averaged over one orbital time (more specifically, between $0.2$ and 
$1.2$ orbits) are
$\left<F_{DF,z}\right>=-0.25$ and $\left<F_{DF,\phi}\right>=-0.18$ in dimensionless units.
These values are very similar to those found in the face-on case.

For $\mu_{\rm cm}=0.5$ and $\mu_{\rm orb}=2$, $F_{DF,z}$ takes positive values between
$2.4$ ($\phi_{p}=144^{\circ}$) and $2.8$ orbits ($\phi_{p}=290^{\circ}$). In that range $\dot{z}_{p}<0$.
The most negative value of $F_{DF,z}$ occurs at $3.12$ orbits ($\phi_{p}=43^{\circ}$). Averaged over
one orbit, $\left<F_{DF,z}\right>=-0.06$. On the other hand, $F_{DF,\phi}$ is always negative,
that is, extracts angular momentum from the binary, and varies between $-1.1$ to $-0.5$, having
$\left<F_{DF,\phi}\right>=-0.7$. This value is similar to the corresponding value in the face-on
orbit.

If $\mu_{\rm cm}$ is significantly greater than $\mu_{\rm orb}$ then $\dot{z}_{p}>0$  and
$F_{DF,z}$ is expected to be always negative. This occurs for $\mu_{\rm cm}=2$ and 
$\mu_{\rm orb}=0.5$.
For those Mach numbers, we find that $\left<F_{DF,z}\right>=-1.18$, again very similar
to the value inferred for the face-on orbit.
On the other hand, $F_{DF,\phi}$ reaches a maximum positive value at $0.5$ orbits
($\phi_{p}=180^{\circ}$) and
then declines to negative values between $0.75$ and $1.2$ orbits. 
The average value of $F_{DF,\phi}$ is close to zero.

Finally, we consider a situation where $\mu_{\rm cm}=\mu_{\rm orb}=1.2$. We see that
$F_{DF,z}$ is always negative and its absolute value increases slowly beyond $1.6$ orbits. At $\tilde{t}=9$, 
$F_{DF,z}$ is about $30\%$ smaller than it is in the face-on case.
The azimuthal component of the drag force is positive between $1.35$ and $1.65$ orbits
($\phi_{p}$ between $125^{\circ}$ and $165^{\circ}$), and negative otherwise.
The average value of $F_{DF,\phi}$ between $1.3$ and $2.3$ orbits is $-0.6$, slightly smaller
than the corresponding face-on value.

\section{Drag force and torque on an equal-mass binary. Face-on case}
\label{sec:equalmass}
\subsection{Binaries with $\mu_{\rm cm}\neq 0$}
In the last Section, we computed the force acting on a binary component due to the gravitational
interaction with its own wake.
Here, we wish to estimate (1) the retarding force responsible
to decelerate the center of mass of the binary, and (2)
the total braking torque on the binary.
We shall concentrate on an equal-mass binary
in face-on orbit since it represents the simplest situation. We also assume that the two
components are on circular orbits with radius $R_{p}$ about their mutual center of mass.

In the face-on case, the binary is subject to a braking force $2F_{DF,z}^{(1)}$
that decelerates the binary as a whole. 
Remind that $F_{DF,z}^{(1)}$ is the vertical component of the total force acting on the
particle $1$, including the contribution of the companion's wake.
We have computed ${\mathcal{R}}_{z}$
defined as the factor that connects $F_{DF,z}^{(1)}$ with $F_{DF,z}$: 
$F_{DF,z}^{(1)}={\mathcal{R}}_{z}F_{DF,z}$. For $\mu_{\rm cm}=0.5$
and $\mu_{\rm orb}=0.5$, we found numerically that ${\mathcal{R}}_{z}=1.7$. For $\mu_{\rm cm}>1$,
${\mathcal{R}}_{z}$ depends on $\tilde{t}$ and on $r_{\rm min}$. At $\tilde{t}=9$, 
our computations indicate that ${\mathcal{R}}_{z}=1.4-1.6$
for $\mu_{\rm cm}$ between $1.2$ and $4$, using $r_{\rm min}=0.1 R_{p}$.
 When $t\rightarrow \infty$, ${\mathcal{R}}_{z}\rightarrow 2$ because the $z$-component
 of the dynamical friction force is dominated by
the far-field wake and, at distances much larger than the binary separation,
the far-field wake is almost identical to the wake produced by a point-like object of mass $2M$.

The braking torque is given by:
\begin{equation}
\Gamma=2R_{p}F_{DF,\phi}^{(1)} .
\end{equation}
We may write
$F_{DF,\phi}^{(1)}={\mathcal{R}}_{\phi}F_{DF,\phi}$. Values ${\mathcal{R}}_{\phi}<1$ indicate
that the wake of the companion reduces the azimuthal drag force. ${\mathcal{R}}_{\phi}$ depends
on $\mu_{\rm cm}$, $\mu_{\rm orb}$ and $r_{\rm min}$. An exhaustive exploration of the dependence
of ${\mathcal{R}}_{\phi}$ on these three parameters is beyond the scope of this paper. Roughly,
we found that for $r_{\rm min}=0.1 R_{p}$ and $\mu_{\rm cm}>1$ or $\mu_{\rm orb}>2$, ${\mathcal{R}}_{\phi}$
lies between $0.57$ and $0.7$. For $\mu_{\rm cm}<1$ and $\mu_{\rm orb}<2$, ${\mathcal{R}}_{\phi}$ takes
somewhat smaller values. For instance, for $\mu_{\rm cm}=0.5$ and $\mu_{\rm orb}=0.5$,
${\mathcal{R}}_{\phi}=0.38$. The ${\mathcal{R}}_{\phi}$-value continues decreasing when 
$\mu_{\rm cm}\rightarrow 0$
and $\mu_{\rm orb}\rightarrow 0$.

Once $F_{DF,\phi}^{(1)}$ is known, it is possible to derive the temporal evolution of
the binary separation.
If each component of the binary has mass $M$,
the total orbital angular momentum is $L=\sqrt{GMR_{p}}$.
Thus, $dL/dt=\Gamma$ implies

\begin{equation}
\frac{dR_{p}}{dt}=2 \sqrt{\frac{R_{p}}{GM}}\Gamma =4\sqrt{\frac{R_{p}^{3}}{GM}}{\mathcal{R}}_{\phi}F_{DF,\phi}.
\label{eq:dR_dt}
\end{equation}
We must just note that we have computed  ${\mathcal{R}}_{\phi}$
and $F_{DF,\phi}$ as a function of $\mu_{\rm cm}$ and
$\mu_{\rm orb}$.  In order to integrate Eq. (\ref{eq:dR_dt}), we may use
that  $\mu_{\rm orb}$ is related to $R_{p}$ through
\begin{equation}
\mu_{\rm orb}=\frac{1}{2c_{\infty}}\sqrt{\frac{GM}{R_{p}}}.
\end{equation}
In the hypothetical case that ${\mathcal{R}}_{\phi}\simeq $const and 
$|F_{DF,\phi}|\propto \mu_{\rm orb}^{-2}$
then $R_{p}(t)=R_{p,0}(1+t/t_{ch})^{-2/3}$, where $t_{ch}$ is a characteristic timescale and $R_{p,0}$ is
the initial radius.

\begin{figure}
\epsfig{file=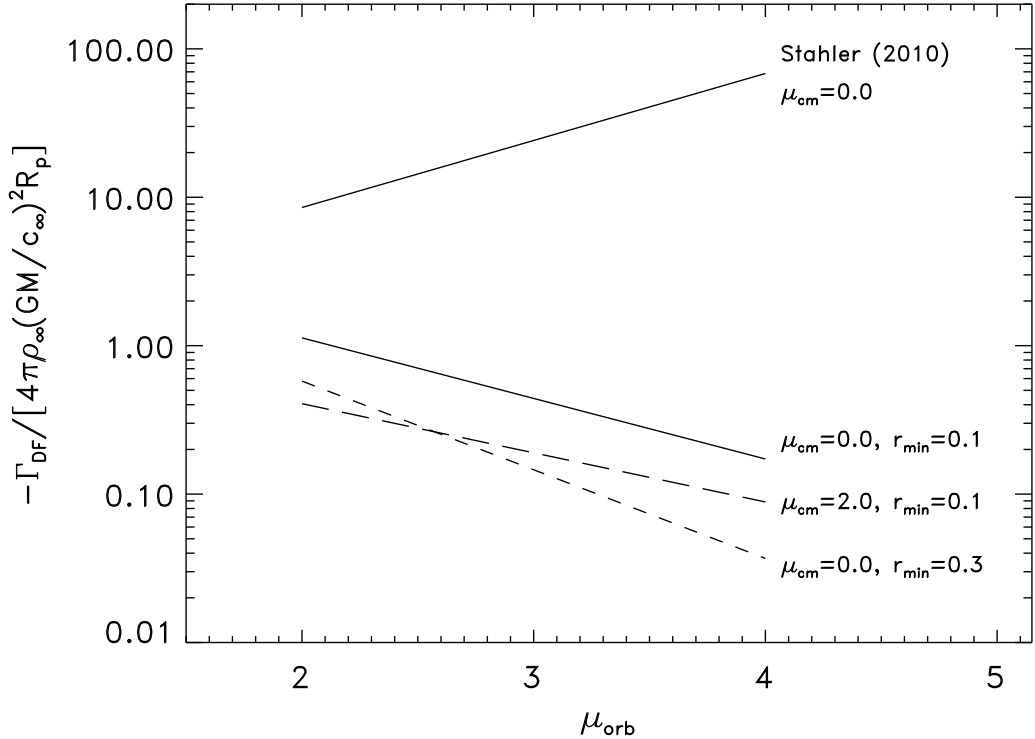,angle=0,width=14.5cm}
  \caption{Comparison of the torque $\Gamma$ on an equal-mass binary as a function of $\mu_{\rm orb}$
using Stahler's formula and our semianalytical scheme. }
\label{fig:torque_Mach}
\vskip 0.75cm
\end{figure}

\subsection{Binaries with $\mu_{\rm cm}=0$. Comparison with previous work}
The torque exerted on a binary 
embedded in a gas was first calculated by Kim et al. (2008), for both
subsonic and supersonic orbital motions, assuming that its center of mass is at rest and the
two stars are on circular orbits.  As already
said,  we were able to reproduce Kim et al. (2008) results from scratch. 

Stahler (2010) also derived analytically the total torque $\Gamma$ on a ``hard'' binary (that is, the separation
of the binary being smaller than the Bondi radius) in the linear approximation theory. 
The gravitational potential
generated by the binary was expanded in multipoles. In the quadrupole approximation, he found
\begin{equation}
\Gamma = -\frac{16\pi}{15} \frac{\Omega^{3}}{c_{\infty}^{5}}\rho_{\infty}G^{2}I^{2},
\end{equation}
where $I$ is the binary's moment of inertia.
For equal-mass binaries, $I=2MR_{p}^{2}$ and $\Omega=V_{p}/R_{p}$. Thus,
\begin{equation}
\Gamma = -\frac{64\pi}{15} \rho_{\infty}\left(\frac{2GM}{c_{\infty}}\right)^{2}R_{p}\mu_{\rm orb}^{3}.
\end{equation}
This cubic dependence of $\Gamma$ with $\mu_{\rm orb}$ is intringuing, especially when the binary
is hard and thereby the relative speed of the stars is supersonic respect to the gas.
This is in sharp contrast with our findings that the torque decreases with $\mu_{\rm orb}$ 
in the supersonic case (also Kim \& Kim 2007).
Figure \ref{fig:torque_Mach} 
compares the net torque on an equal-mass binary using our approach (which is identical, but
independent, to that used by Kim et al. 2008) and using Stahler (2010) formula.
We find clear and significant disagreement with the Stahler (2010) work.
For $\mu_{\rm orb}\simeq 4$,
the latter formula predicts a torque two orders of magnitude larger than what we got
even assuming a rather small value of $0.1R_{p}$ for $r_{\rm min}$.
Note that Stahler's formula has not any explicit dependence on $r_{\rm min}$.

\begin{figure}
\epsfig{file=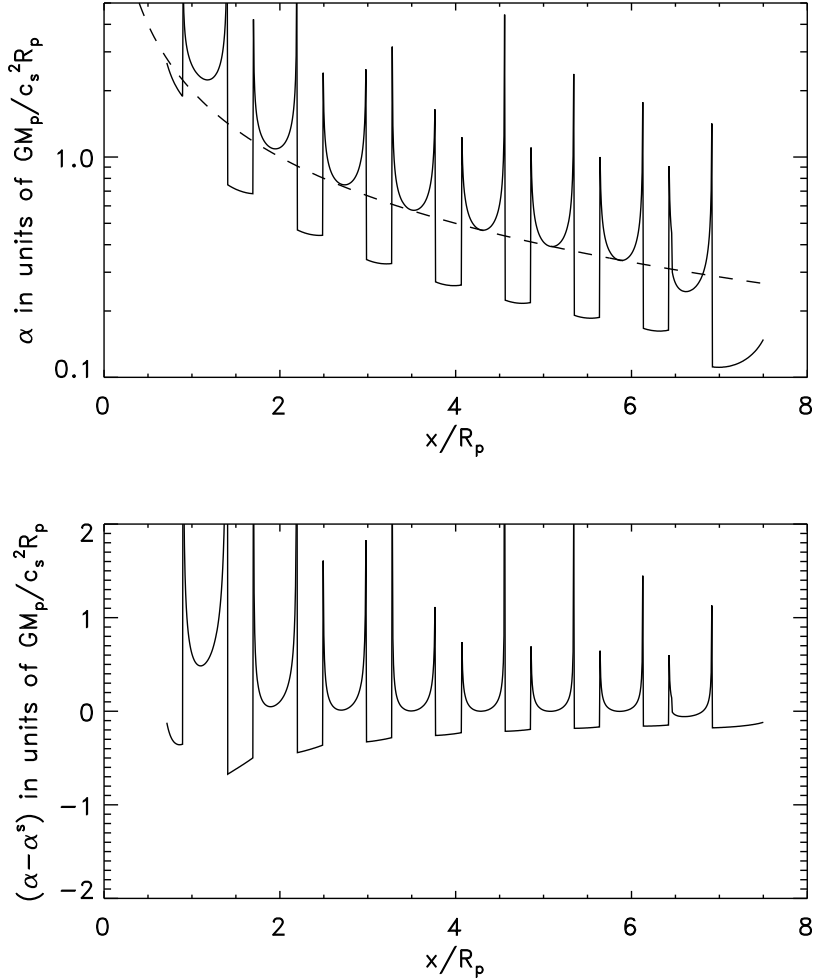,angle=0,width=12.0cm}
  \caption{Top panel: Density enhancement $\alpha$ (solid line) and monopole contribution
$\alpha_{\mbox{\tiny -1}}^{s}$ (dashed line)
along the positive $x$-axis for an equal-mass binary rotating at $\mu_{\rm orb}=4$
and $\mu_{\rm cm}=0$, at $\tilde{t}=7.5$. Bottom panel: Distribution of $\alpha-\alpha_{\mbox{\tiny -1}}^{s}$.
As the orbital motion is supersonic, the distributions exhibit sharp discontinuities. }
\label{fig:rho_cut_double}
\vskip 0.75cm
\end{figure}

Stahler (2010) carried out a multipole expansion of the gravitational potential created by the
binary. In this approach, the perturbed density of the gas can be decomposed as
\begin{equation}
\alpha = \alpha_{\mbox{\tiny -1}}+\alpha_{\mbox{\tiny -2}}+\alpha_{\mbox{\tiny -3}}+...,
\end{equation}
where $\alpha_{\mbox{\tiny -1}}$ is a term that decays as $r^{-1}$, 
$\alpha_{\mbox{\tiny -2}}$ decays as $r^{-2}$ and so on.
The perturbed density $\alpha_{\mbox{\tiny -1}}$ according to Stahler (2010) is
\begin{equation}
\alpha_{\mbox{\tiny -1}} = \frac{GM_{\rm tot}}{c_{\infty}^{2}r}-\frac{GI\Omega^{2}}{c_{\infty}^{4}r}\sin^{2}\theta \cos 2\left(\Omega t-\frac{\Omega}{c_{\infty}}r-\phi\right)+{\rm higher \, \, order \,\, terms},
\label{eq:stahler1}
\end{equation}
where $M_{\rm tot}$ is the total mass of the binary and $\theta$ and
$\phi$ are the polar and azimuthal angles, respectively.
The first term on the right-hand-side of Equation (\ref{eq:stahler1}) corresponds to the monopole contribution 
of the binary, which produces a
spherical and static envelope. We will denote it by $\alpha_{\mbox{\tiny -1}}^{s}$. The second term
on the right-hand-side of Equation (\ref{eq:stahler1}), which
we will refer to it as $\alpha_{\mbox{\tiny -1}}^{\rm osc}$, is produced by the quadrupole portion.

The smooth behaviour of the perturbed density distribution described by  
Equation (\ref{eq:stahler1}), which
oscillates sinusoidally, is in disagreement with the density profile found in our supersonic calculations.
As an example, Figure \ref{fig:rho_cut_double} shows the dimensionless perturbed density $\alpha$, as inferred in our
mathematical scheme, through a cut
along the positive $x$-axis for $\mu_{\rm orb}=4$ (and $\mu_{\rm cm}=0$), for an equal-mass
binary (at $\tilde{t}=7.5$). We include the wake induced by both components of the binary.
The dimensionless static density $\alpha^{s}_{\mbox{\tiny -1}}$ is also drawn. Interestingly, 
$\alpha^{s}_{\mbox{\tiny -1}}$ traces the density at the center of the
spiral arms. We see that, in our computations, $\alpha-\alpha^{s}_{\mbox{\tiny -1}}$ oscillates about zero,
but not at all in a sinusoidal way.

The source of the disagreement between our results and Stahler (2010) can be traced back
to the multipole expansion, which is not a suitable choice to solve the inhomogenous
wave equation (\ref{eq:inhomogeneous_equation}).
Let us rewrite Eq.~(\ref{eq:stahler1}) in terms of $r_{\rm in}\equiv GM_{\rm tot}/(2c_{\infty}^{2})$ 
and $\Omega_{s}\equiv c_{\infty}/R_{p}$.  For equal-mass binaries:
\begin{equation}
\alpha_{\mbox{\tiny -1}}= \frac{2r_{\rm in}}{r}\left[1-\left(\frac{\Omega}{\Omega_{s}}\right)^{2}+
{\mathcal{O}}\left(\frac{\Omega^{4}}{\Omega_{s}^{4}}\right)\right].
\end{equation}
This implies that the multipole expansion is equivalent to a series expansion in powers
of $\Omega/\Omega_{s}$. Convergence would require $\Omega\ll \Omega_{s}$, which
implies $\mu_{\rm orb}\ll 1$. Thus, the multipole approach is not valid, at least, for 
hard binaries. We must warn that the formal solution of the inhomogeneous wave equation
given in Equation (\ref{eq:formal_solution}) is an integral over the whole volume. Thus, in order
to perform this integral, it is desirable to have a good approximation of
$\rho_{\rm ext}$ at every point and not only in the far field, even if $\mu_{\rm orb}\ll 1$.

Korntreff et al. (2012) used Stahler's formula to model the process of the gas-induced orbital decay 
in hard binaries. They found that orbital decay due to gas-damping can reshape the period
distribution of short-period binaries. This conclusion should be revised using more accurate
estimates for the torque and including the fact that binaries are in orbits around the center of
the star-forming clusters.

\section{Summary}
\label{sec:conclusions}

There is a variety of astrophysical scenarios where a gravitationally bound system
comprised of small gravitational objects, moves relative to the ambient medium. 
For example, a globular cluster embedded in the gaseous halo of a protogalaxy,
a star cluster moving through the disk of a gas-rich galaxy, a binary star traveling 
within its natal gas cloud, or a binary star embedded in a central stellar cluster subject
to a major gas inflow.
We have considered the dynamical friction in the context of a binary system but our analysis 
may be extended to study other astrophysical situations where the perturbers do not move 
on rectilinear orbits. In particular, we have provided the mathematical framework to
study the gravitational interaction of a binary system with its surrounding gas, when
it moves through a uniform, static gaseous background, in linear theory. We have shown how 
the presence of a binary (instead of a single point-mass) and of the additional component 
center-of-mass velocity modifies the morphology of the density wake and the nature of the 
drag force. We have also quantified the internal torques.

Using time-dependent linear perturbation theory, we have developed a semi-analytical
scheme to derive the enhanced-density wake created by a binary system.
The method is an extension of the work of Kim \& Kim (2007).
We have first examined the gravitational wake excited in the medium by just
one component of the binary. 
This facilitates comparison with previous studies that consider one single
perturber. For simplicity, we have assumed that it is on a circular orbit about
the center of mass of the binary.
A model is specified with three dimensionless parameters: $\mu_{\rm cm}$ 
(the Mach number associated with the center of mass), $\mu_{\rm orb}$ (the Mach
number of the orbital motion) and $i$ the angle between the orbital axis and $\VV_{\rm cm}$.
We have characterized the wake and computed the drag forces in two cases: 
when the orbital plane is perpendicular to $\VV_{\rm cm}$ (face-on case)
and when they are parallel (edge-on case).

The morphology of the wake depends on whether the motions are supersonic or subsonic.
If both are subsonic (i.e. $\mu_{\rm cm}<1$ and $\mu_{\rm orb}<1$), the perturbed
density is confined to the sonic sphere centered in the initial position of the 
perturber. Within the sonic sphere, the perturbed density distribution  displays
a smooth comma-shape tail along cuts parallel and perpendicular to the orbital plane.
When $\mu_{\rm cm}<1$ and $\mu_{\rm orb}>1$, the perturber torques the gas
and launches spiral trailing waves which transport angular momentum.
Finally,  when $\mu_{\rm cm}>1$ and $\mu_{\rm orb}>1$,
the wake displays a very complicated structure, unless $\mu_{\rm cm}$ and $\mu_{\rm orb}$ 
are very different in magnitude.
If $\mu_{\rm cm}\gg\mu_{\rm orb}$,
the perturbed density distribution is confined within a deformed Mach cone.

We have computed the 
drag forces that arise from the gravitational attraction between the perturber and the wake
it excites in the ambient medium.  
We have compared the drag force in the $z$-direction with that prediced using the Ostriker formula 
which was derived for a body in rectilinear orbit.  If the perturber moves at  $\mu_{\rm cm}>1$, 
the dynamical friction force in the $z$-direction is reduced as compared to the rectilinear
case because of the loss of gravitational focusing of gas behind the perturber, due
to the orbital motion. However, if $\mu_{\rm cm}<1$, the $z$-component of the drag
force may be larger than in the pure rectilinear case for certain values of $\mu_{\rm orb}$.
For instance, for $i=0$, $\mu_{\rm cm}=0.5$, $\mu_{\rm orb}=1.2$ and $r_{\rm min}=0.1R_{p}$, 
the $z$-component
of the drag force is several times larger than it is in the pure rectilinear case ($\mu_{\rm orb}=0$).

We have also investigated the azimuthal component of the drag force, which causes
the binary to lose orbital angular momentum. We have found that Kim \& Kim (2007) formula
provides reasonable estimates of $F_{DF,\phi}$ as long as $\mu_{\rm cm}<0.5$ or 
$\mu_{\rm orb}\gg\mu_{\rm cm}$.
In the face-on case, we find that,
at a fixed value of $\mu_{\rm orb}$ larger than $1$, $F_{DF,\phi}$ decreases
with increasing  $\mu_{\rm cm}$. As illustration, for $\mu_{\rm orb}=1.2$,
$F_{DF,\phi}$ is a factor of $30$ smaller for $\mu_{\rm cm}=4$ than it is
for $\mu_{\rm cm}=0$.
However, for $\mu_{\rm orb}<1$, $F_{DF,\phi}$
versus $\mu_{\rm cm}$ presents a maximum around $\mu_{\rm cm}\approx 1$.
Therefore, it is misleading to think that the translational motion of the center of mass
always leads to a decrease in $F_{DF,\phi}$.

Our analysis is especially relevant to understand the gas-induced effects
on binary stars when they are still embedded in their parent cloud, or in a central stellar cluster
that is undergoing a major inflow of gas (e.g., Davies et al. 2011). 
Dynamical friction  may cause  
orbital migration of the binary towards the center of the cloud, leading to mass
segregation. In addition, the binary can shrink by the tidal torque created by
its own wake, reducing its orbital period. 
We have compared the braking torque acting on an equal-mass binary
at rest relative to the ambient medium with that using the analytical formula derived by Stahler (2010).
We find that Stahler's formula overestimates the torque. For instance, for $\mu_{\rm orb}=4$,
Stahler's formula overestimates the torque by two orders of magnitude.
We conclude that the analysis made by Korntreff et al. (2012) should be revised using the
correct magnitude for the torques and including the translational motion of the binaries.

\section*{acknowledgements}
We would like to thank Ana Hidalgo G\'amez and Gonzalo Ares de Parga for useful discussions and
for their advice in many aspects. Our special thanks to Nathan Leigh and our anonymous
referee for constructive comments on the manuscript. This work was partly supported
by CONACyT project 165584 and PAPIIT project IN106212. Ra\'ul O. Chametla acknowledges
financial support from a CONACyT scholarship and the grant BEIFI 20141185.

\appendix
\section{Face-on case}
\label{sec:app_faceon}
 Using cylindrical coordinates $(R, \phi, z)$, where $R$
is the cylindrical radius and $\phi$ the azimuthal angle,
Equation (\ref{eq:D_general}), with $i=0$, is reduced to
\begin{equation}
{\mathcal{D}}(\rr,t)=\sum_{\varphi_{j}}\frac{\mu_{\rm orb}}{|(1-\mu_{\rm cm}^{2})\varphi_{j}-\Omega t-\mu_{\rm orb}^{2}
[\tilde{R} \sin(\varphi_{j}-\phi)-\lambda\tilde{z}]|}\hcur\left(\frac{\varphi_{j}}{\Omega}\right).
\end{equation} 
Here $\varphi_{j}$ are the solutions of the following equation:
\begin{equation}
\mu_{\rm orb}\tilde{d}=-(\varphi-\Omega t),
\end{equation}
where
\begin{equation}
\tilde{d}(\varphi;\rr)=[(\tilde{x}-\cos\varphi)^{2}+(\tilde{y}-\sin \varphi)^{2}
+(\tilde{z}-\lambda\varphi)^{2}]^{1/2}=
[1+\tilde{R}^{2}-2\tilde{R}\cos(\varphi-\phi)
+(\tilde{z}-\lambda\varphi)^{2}]^{1/2}.
\end{equation}
If we define $\omega\equiv\varphi-\phi$, $\eta\equiv\phi-\Omega t$ and $\tilde{a}=(z-V_{\rm cm}t)/R_{p}$,
then
\begin{equation}
{\mathcal{D}}(\rr,t)=\sum_{\omega_{j}}\frac{\mu_{\rm orb}}{|(1-\mu_{\rm cm}^{2})(\omega_{j}+\eta)-\mu_{\rm orb}^{2}
\tilde{R} \sin\omega_{j}+\mu_{\rm orb}\mu_{\rm cm}\tilde{a}|}
\hcur\left(\frac{\omega_{j}+\phi}{\Omega}\right),
\label{eq:Dfaceon}
\end{equation} 
where $\omega_{j}$ are the solutions of the equation
\begin{equation}
\mu_{\rm orb}\tilde{d}_{j}=-(\omega_{j}+\eta),
\label{eq:eqfaceon}
\end{equation}
with
\begin{equation}
\tilde{d}_{j}(\tilde{R},\tilde{a},\eta)=[1+\tilde{R}^{2}-2\tilde{R}\cos \omega_{j}
+(\tilde{a}-\lambda[\omega_{j}+\eta])^{2}]^{1/2}.
\label{eq:rootsfaceon}
\end{equation}

The perturbed density distribution created  by a perturber in pure circular motion was studied 
by Kim \& Kim (2007). This case correponds
to $i=0$ and $\mu_{\rm cm}=0$. In that situation, Equations (\ref{eq:Dfaceon}) and (\ref{eq:rootsfaceon})
simplify to
\begin{equation}
{\mathcal{D}}(\rr,t)=\sum_{\omega_{j}}\frac{\mu_{\rm orb}}{|\omega_{j}+\eta-\mu_{\rm orb}^{2}
\tilde{R} \sin\omega_{j}|}
\hcur\left(\frac{\omega_{j}+\phi}{\Omega}\right),
\label{eq:DKK}
\end{equation}
and 
\begin{equation}
\tilde{d}_{j}(\tilde{R},\tilde{z})=(1+\tilde{R}^{2}-2\tilde{R}\cos \omega_{j}
+\tilde{z}^{2})^{1/2},
\end{equation}
which correspond to Eqs (8) and (6), respectively, in Kim \& Kim (2007); just note that they used the letter
$s$ instead of $\eta$.

\section{Edge-on case}
\label{sec:app_edgeon}
The edge-on case corresponds to $i=\pi/2$. For this value of the inclination angle, Equation
(\ref{eq:D_general}) is simplified to 
\begin{equation}
{\mathcal{D}}(\rr,t)=\sum_{\phi_{j}}\frac{\mu_{\rm orb}}{|(1-\mu^{2}_{\rm cm})\varphi_{j}-\Omega
t-\mu_{\rm orb}^{2}
[\tilde{x}\sin\varphi_{j}-\tilde{z}(\cos\varphi_{j}+\lambda)+h(\varphi_{j})]|}
\hcur\left(\frac{\varphi_{j}}{\Omega}\right).
\end{equation} 
 The $\varphi_{j}$-values are solutions of
\begin{equation}
\mu_{\rm orb}\tilde{d}_{j}=-(\varphi_{j}-\Omega t),
\end{equation}
where
\begin{equation}
\tilde{d}_{j}(\rr)=[(\tilde{x}-\cos\varphi_{j})^{2}+\tilde{y}^{2}
+(\tilde{z}-\sin\varphi_{j}-\lambda\varphi_{j})^{2}]^{1/2}
\end{equation}
Recall that $\Omega=\mu_{\rm orb}$ in dimensionless units made with $R_{p}$ and $c_{\infty}$.

\section{The location of the roots}
\label{sec:app_location}
In order to not miss any root, we need to constraint the interval that contains the roots.
For illustration, we outline the procedure for the case when the orbit is circular
and face-on.  According to Eqs. (\ref{eq:eqfaceon}) and (\ref{eq:rootsfaceon}),
the equation for the roots is
\begin{equation}
\mu_{\rm orb}[1+\tilde{R}^{2}-2\tilde{R}\cos \omega_{j}
+(\tilde{a}-\lambda[\omega_{j}+\eta])^{2}]^{1/2}=-(\omega_{j}+\eta).
\label{eq:rootsequation}
\end{equation}
The left-hand-side of this equation is positive (or zero), thus $-(\omega_{j}+\eta)$ must be
positive (or zero); this implies that the roots are in the range $\omega\leq-\eta$.

Now we search for a lower bound of the interval.
We know that $1+\tilde{R}^{2}-2R\cos \omega \geq (\tilde{R}-1)^{2}$, hence
\begin{equation}
\mu_{\rm orb}^{2}[1+\tilde{R}^{2}-2\tilde{R}\cos \omega
+(\tilde{a}-\lambda[\omega+\eta])^{2}]\geq
\mu_{\rm orb}^{2}[(\tilde{R}-1)^{2}
+(\tilde{a}-\lambda[\omega+\eta])^{2}].
\end{equation}
Therefore, it is obvious that those $\omega$'s satisfying the inequality
\begin{equation}
\mu_{\rm orb}^{2}[(\tilde{R}-1)^{2}
+(\tilde{a}-\lambda[\omega+\eta])^{2}]> (\omega+\eta)^{2},
\end{equation}
cannot be roots of Equation (\ref{eq:rootsequation}).
This inequatity can be written as
\begin{equation}
(\mu_{\rm cm}^{2}-1)(\omega+\eta)-2\mu_{\rm cm}\mu_{\rm orb}\tilde{a}(\omega+\eta)
+\mu_{\rm orb}^{2}([\tilde{R}-1]^{2}+\tilde{a}^{2})>0.
\end{equation}
Now suppose that 
$\mu_{\rm cm}>1$, the above inequality implies that, if there are any roots of Equation 
(\ref{eq:rootsequation}),
they should be located in the interval $\omega_{-}\leq \omega\leq \omega_{+}$ with
\begin{equation}
\omega_{\pm}={\rm min}\left[0,\frac{\mu_{\rm orb}}{\mu_{\rm cm}^{2}-1}\left(\mu_{\rm cm}\tilde{a}\pm\sqrt{\tilde{a}^{2}-(\mu_{\rm cm}^{2}-1)(\tilde{R}-1)^{2}}\right)\right]-\eta.
\end{equation}
We see that when $\tilde{a}^{2}< (\mu_{\rm cm}^2-1)(\tilde{R}-1)^{2}$ or 
$\mu_{\rm cm}\tilde{a}-\sqrt{\tilde{a}^{2}-(\mu_{\rm cm}^{2}-1)(\tilde{R}-1)^{2}}>0$, 
there is no roots.
If $\tilde{a}>0$, the second inequality is
$\mu_{\rm cm}^2\tilde{a}^{2}>\tilde{a}^{2}-(\mu_{\rm cm}^{2}-1)(\tilde{R}-1)^{2}$, or
$(\mu_{\rm cm}^{2}-1)\tilde{a}^{2}>-(\mu_{\rm cm}^{2}-1)(\tilde{R}-1)^{2}$, which always
holds. Therefore, the second inequality implies that $\tilde{a}\leq 0$ is a necessary
condition (not sufficient) to have roots (note
that we are studying the case $\mu_{\rm cm}>1$). Putting together the first and
the second inequalities, we obtain that the density perturbation is confined to
a region having
$\tilde{a}\leq 0$ and 
$\tilde{a}\geq -(\mu^{2}_{\rm cm}-1)^{1/2}(\tilde{R}-1)$. We refer to this region 
as the ``large-scale rear Mach cone'', which was mentioned in Section \ref{sec:wakefaceon}.

In order to find the interval that contains the roots when $\mu_{\rm cm}<1$,
it is useful to start from the following inequality:
\begin{equation}
1+\tilde{R}^{2}-2\tilde{R}\cos \omega
+(\tilde{a}-\lambda[\omega+\eta])^{2}\leq
(\tilde{R}+1)^{2}
+(\tilde{a}-\lambda[\omega+\eta])^{2}.
\end{equation}
Therefore, those $\omega$'s satisfying the condition
\begin{equation}
\mu_{\rm orb}^{2}[(\tilde{R}+1)^{2}
+(\tilde{a}-\lambda[\omega+\eta])^{2}]< (\omega+\eta)^{2},
\end{equation}
cannot be roots.
After some simple algebra,
one can determine the bounded interval $\omega_{-}\leq \omega \leq \omega_{+}$ where the roots 
should lie (when $\mu_{\rm cm}<1$). We find that 
$\omega_{+}=-\eta$ and
\begin{equation}
\omega_{-}=-\frac{\mu_{\rm orb}}{1-\mu_{\rm cm}^{2}}\left(\mu_{\rm cm}\tilde{a}+\sqrt{\tilde{a}^{2}+(1-\mu_{\rm cm}^{2})(\tilde{R}+1)^{2}}\right)-\eta.
\end{equation}
Contrary to the case $\mu_{\rm cm}>1$,   the interval is never null if $\mu_{\rm cm}<1$.
In fact, it is easy to show that there is always at least one root for $\mu_{\rm cm}<1$. 

We have carried out a similar analysis to constrain the interval to search for the roots
in all the cases presented in this paper.

\end{document}